%% file: main.tex
\UseRawInputEncoding
\documentclass[sigconf]{styles/acmart}
\PassOptionsToPackage{table}{xcolor} 

\theoremstyle{acmplain}
\newtheorem{theorem}{Theorem}
\theoremstyle{acmdefinition}
\newtheorem{definition}{Definition}

\AtBeginDocument{%
  }

\setcopyright{acmlicensed}
\copyrightyear{2026}\acmYear{2026}
\setcopyright{cc}
\setcctype{by}
\acmConference[CPSIoTSec '26]{8th Workshop on CPS\&IoT Security and Privacy}{November 15-19, 2026}{The Hague, Netherlands}
\acmBooktitle{8th Workshop on CPS\&IoT Security and Privacy (CPSIoTSec '26), November 15-19, 2026, The Hague, Netherlands}
\acmDOI{10.1145/3847353.3847504}
\acmISBN{979-8-4007-3032-0/2026/11}

\usepackage{listings}
\lstdefinestyle{terminal}{backgroundcolor=\color{black!5}, basicstyle=\ttfamily\small, keywordstyle=\color{blue}, commentstyle=\color{gray}, stringstyle=\color{red}}
\usepackage{subcaption}
\usepackage{booktabs}
\usepackage[table]{xcolor}
\usepackage{float}
\usepackage{graphicx}
\usepackage{svg}
\usepackage{adjustbox}
\usepackage{cleveref}
\usepackage[acronym]{glossaries} 
\usepackage{multicol}
\usepackage{balance}

\newif\ifshowcomments

\input{preamble}

\usepackage{xcolor}

\newcounter{reviewer}
\newcounter{comment}[reviewer]

\begin{document}

\thispagestyle{plain}
\newpage

\setcounter{page}{1}
\twocolumn


\newacronym{CPS}{CPS}{Cyber-Physical Systems}
\newacronym{IDS}{IDS}{Intrusion Detection System}
\newacronym{FDIR}{FDIR}{Fault Detection, Isolation, and Reconfiguration}
\newacronym{ROS2}{ROS2}{Robot Operating System 2}
\newacronym{Webots}{Webots}{Webots Simulator}
\newacronym{PR2}{PR2}{Personal Robot 2}
\newacronym{ARCANE}{ARCANE}{Adaptive Resilience for Cyber-Attacks in Node-level Embodiments}
\newacronym{RobResilience}{RobResilience}{Implementing and Evaluating a Resilience Framework for Cyber-Physical Embodied Systems}

\title{RobResilience: Implementing and Evaluating \\ a Resilience Framework for Cyber-Physical Embodied Systems}


\author{Gysella Imrell}
\affiliation{%
  \institution{\"Orebro University}
  \city{\"Orebro}
  \country{Sweden}
}
\email{gysella.michelle@gmail.com}

\author{Emanuele Miotto}
\affiliation{%
  \institution{\"Orebro University}
  \city{\"Orebro}
  \country{Sweden}}
\email{emanuele.miotto@oru.se}

\author{Mahya Mohammadi Kashani}
\correspondingauthor
\affiliation{%
  \institution{\"Orebro University}
  \city{\"Orebro}
  \country{Sweden}}
\email{mahya.mohammadi-kashani@oru.se}

\author{Mauro Conti}
\affiliation{%
  \institution{\"Orebro University}
  \city{\"Orebro}
  \country{Sweden}}
\affiliation{%
  \institution{University of Padua}
  \city{Padua}
  \country{Italy}}
\email{mauro.conti@unipd.it}

\author{Alberto Giaretta}
\affiliation{%
  \institution{\"Orebro University}
  \city{\"Orebro}
  \country{Sweden}}
  \email{alberto.giaretta@oru.se}

\renewcommand{\shortauthors}{Imrell et al.}

\input{chapters/abstract}





\maketitle

\input{chapters/introduction}
\input{chapters/related_work}
\input{chapters/resilience_framework}
\input{chapters/threat_model}
\input{chapters/implementation}
\input{chapters/evaluation}
\input{chapters/future_work}
\input{chapters/conclusion}
\input{chapters/ethics}
\input{chapters/genAI_usage}
\input{chapters/acks}


\bibliographystyle{references/ACM-Reference-Format}
\balance
\bibliography{references/references}



\end{document}

%% file: preamble.tex
\makeglossaries

\newacronym{cps}{CPS}{Cyber-Physical System}
\newacronym{cps-pl}{CPSs}{Cyber-Physical Systems} 
\newacronym{ids}{IDS}{Intrusion Detection System}
\newacronym{ros2}{ROS2}{Robot Operating System 2}
\newacronym{webots}{Webots}{Webots Simulator}
\newacronym{pr2}{PR2}{Personal Robot 2}
\newacronym{fdir}{FDIR}{Fault Detection, Isolation, and Reconfiguration}
\newacronym{arcane}{ARCANE}{Adaptive Resilience for Cyber-Attacks in Node-level Embodiments}
\newacronym{robresilience}{RobResilience}{Implementing and Evaluating a Resilience Framework for Cyber-Physical Embodied Systems}
\newacronym{dos}{DoS}{Denial-of-Service}


%% file: chapters/abstract.tex
\begin{abstract}
  In embodied cyber-physical systems, active cyberattacks pose an immediate threat not just to data, but to physical integrity and human safety. While existing security approaches excel at detection, they lack the runtime mechanisms to determine whether a disruption is tolerable or if performance degradation remains within safe operational bounds. This gap leaves autonomous systems vulnerable to graceful failure paralysis, where they cannot distinguish between a safe, degraded state and a catastrophic hazard during an ongoing attack.
  This paper presents RobResilience, an implementation of a formal resilience framework for embodied cyber-physical systems
  in a \gls{webots} simulation environment, using a \gls{pr2} robot and \gls{ros2}.
  The framework evaluates three predicates at runtime: tolerable disruption~($\delta$), tolerable degradation~($\gamma$), and mitigation feasibility~($\mu$), over a compromised device set derived from IDS confidence scores. When resilience is lost, the framework triggers available mitigation strategies. We evaluate our implementation through eight attack scenarios that systematically cover all possible combinations of the predicate state space, varying attack targets, degradation rates, and mitigation availability.  Results confirm that the runtime behaviour of the implementation is consistent with the theoretical definitions. 
  ~\footnote{The full reproducible implementation is available at: \url{https://github.com/mahyamkashani/RobResilience}}\looseness=-1

\end{abstract}

\keywords{cyber-physical systems, device-level cyberattacks, simulation, adaptive resilience framework}

%% file: chapters/introduction.tex
\section{Introduction}
\label{sec:introduction}
\gls{cps-pl} integrate computational and physical components to operate in safety and mission-critical domains, including smart grids, healthcare, autonomous vehicles, and robotics~\cite{karnouskosCyberPhysicalSystemsSmartGrid2011, zhangHealthCPSHealthcareCyberPhysical2017,
siddappajiRoleCyberSecurity2020}. This tight integration extends the attack surface considerably: incidents such as Stuxnet, the Ukrainian power grid blackout, and the Florida water treatment attack have demonstrated that cyberattacks on \gls{cps-pl} can compromise physical
processes, operational continuity, and human safety simultaneously~\cite{kayanCybersecurityIndustrialCyberPhysical2022, duoSurveyCyberAttacks2022}.

A particularly consequential subclass is \emph{embodied} \gls{cps-pl}, autonomous systems that interact with humans and their environment through sensing, decision-making, and physical actuation~\cite{shea-blymyerAlgorithmicEthicsFormalization2021}. In these systems, an attack can affect not only digital assets, but also the physical integrity and operational functionality of the robot and its surroundings, making resilience a first-order concern~\cite{xingRobustSecureEmbodied2026}. 
Cornelius et al. showed that robots like PR2 rely on commonly used Linux operating systems that are difficult to patch, leaving known vulnerabilities unaddressed in deployed systems~\cite{cornelius2017perspective}.
Dudek and Szynkiewicz analysed attack vectors for mobile robots drawing on real-world scenarios spanning network, personal data, services and applications vulnerabilities~\cite{dudek2019cyber}. A major issue with embodied CPSs is that they use AI-powered technology to make autonomous decisions while interacting with people. This creates significant risks to peoples' daily activities if security is not considered. For example, two viral videos on social media in 2016 and similar incidents involving humanoid robots covered by a viral video in 2026 highlight these dangers\cite{fusion2015, incident2026humanoid-robot}.\color{black} Existing work on CPS security has focused mainly on attack detection, intrusion detection, and anomaly analysis~\cite{mitchellSurveyIntrusionDetection2014, hanIntrusionDetectionCyberPhysical2014a,
duoSurveyCyberAttacks2022}. Although valuable for identifying threats, these approaches leave runtime resilience, adaptive mitigation, and graceful degradation comparatively under-explored~\cite{segovia-ferreiraSurveyCyberResilienceApproaches2024, kholidyAutonomousMitigationCyber2021}. Intrusion response work, such as the Autonomous Response Controller of Kholidy, addresses runtime response beyond detection, but does not reason about disruption or degradation tolerability under ongoing attacks~\cite{kholidyAutonomousMitigationCyber2021}.

This paper addresses this gap by implementing and evaluating a formal resilience framework for embodied \gls{cps-pl}~\cite{giarettaFormalResilienceFramework2026}. The framework integrates IDS-derived information with task-criticality assessments, embodiment goals, and device dependencies to reason about disruption tolerability, degradation severity, mitigation feasibility, and operational continuity, going beyond detection to support resilient behaviour during active cyberattacks.

\begin{figure}[t]
  
  \centering 
  \centering
    \includegraphics[width=0.85\columnwidth] %
    {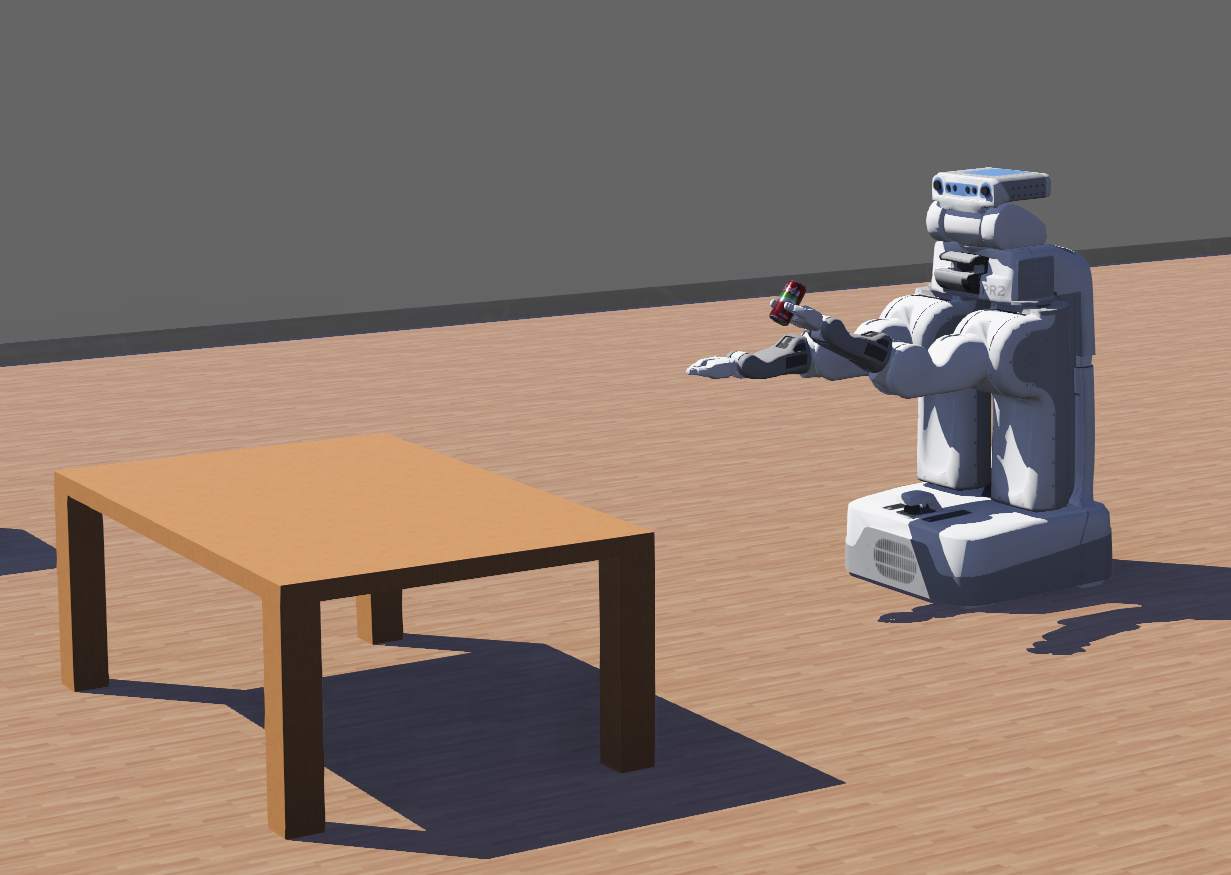}
  \caption{%
    PR2 robot is doing \texttt{navigate\_and\_pickup} task in Webots simulation.%
  }%
  \label{fig:simulation}
\end{figure}

This paper follows a Design Science Research approach, situating the framework within a simulation environment that models how different cyberattack scenarios influence system behaviour. Attacks are simulated as message-based injections over \gls{ros2}~\cite{al2024ros}, targeting a \textit{\gls{pr2}} robot in \gls{webots}~\cite{michel2004cyberbotics}; robot \gls{ids} provides runtime input to the resilience reasoning layer. The \gls{pr2} platform is chosen for its combination of mobility and manipulation capabilities, which enables us to simulate diverse task scenarios, shown in \Cref{fig:simulation}. Experiments are designed to verify consistency with the theoretical framework through the systematic variation of attack targets, degradation rates, and mitigation configurations. Evaluation is limited to simulation and does not extend to physical hardware deployment; in addition, attack scenarios model the expected behavioural effects of cyberattacks on system components rather than real-world exploits. 

\setlength{\textfloatsep}{8pt plus 2pt minus 2pt}
\setlength{\abovecaptionskip}{6pt}

The paper provides three main contributions. First, it presents an executable implementation of the formal resilience framework of~\cite{giarettaFormalResilienceFramework2026} for embodied \gls{cps-pl}, realising the criticality mappings, \gls{ids} integration, and runtime predicate evaluation as deployable software modules within a \gls{ros2}-based \gls{webots} simulation environment. Second, it 
demonstrates that the implementation conforms to the expected behaviour across all realisable combinations among \color{black} the tolerable disruption, tolerable degradation, and mitigation feasibility predicates, confirming consistency with the formal definitions across critical and non-critical attack targets. Third, it delivers an empirical analysis of how degradation severity, task-criticality mappings, and adaptive mitigation jointly determine operational continuity and resilient system behaviour under ongoing cyberattacks.

The remainder of this paper is organised as follows. \Cref{sec:related} reviews related work on CPS security, resilience, and fault detection. \Cref{sec:framework} presents the formal resilience framework, defining the criticality mappings, \gls{ids} model, as well as the predicates for tolerable disruption, tolerable degradation, mitigation feasibility, and system resilience. \Cref{sec:threat-model} describes the threat model, covering attacker goals, capabilities, knowledge assumptions, and defences. \Cref{sec:implementation} details the proposed implementation, including the system architecture, the four framework modules, and the instantiation of the degradation function $\psi$. \Cref{sec:result} presents the evaluation: the experimental setup, eight attack scenarios covering all realisable combinations of the tolerable disruption, tolerable degradation, and mitigation feasibility predicates, and the resulting analysis. \Cref{sec:conclusion} concludes the paper and outlines directions for future work.\looseness=-1

%% file: chapters/related_work.tex
\section{Related Work}
\label{sec:related}
We survey three bodies of work that contextualize \gls{robresilience}.
We first review intrusion detection approaches for \gls{cps-pl} and their limitations in providing adaptive runtime response.
We then examine \gls{fdir} methods
and explain why assumptions about bounded physical faults do not transfer to adversarial compromise.
Finally, we consider formal and runtime-verification approaches to cyber-resilience
for \gls{cps-pl} and embodied systems, identifying the predicate-based framework that \gls{robresilience} implements and evaluates.

\subsection{CPS security and intrusion detection}
\gls{cps-pl} face a diverse and well-documented
attack surface spanning both network and physical
layers~\cite{duoSurveyCyberAttacks2022, kayanCybersecurityIndustrialCyberPhysical2022}.
The dominant response in the literature has been detection. Mitchell and Chen survey intrusion detection techniques for \gls{cps-pl} along two design dimensions, detection technique and audit material, establishing the breadth of
\gls{ids}-centric approaches~\cite{mitchellSurveyIntrusionDetection2014}. Detection techniques fall into three categories. Knowledge-based (misuse) detection matches runtime features against known patterns of bad behaviour; its advantage is a low false-positive rate, but its disadvantage is that it requires a constantly updated dictionary of every known attack vector, so it cannot catch attacks not already specified. Behaviour-based detection instead flags runtime features that deviate from "normal," defined either from the test signal's own history (unsupervised) or from training data (semi-supervised); its advantage is that it does not need attacks pre-specified, but it is prone to false positives and its training/profiling phase is a weakness against sophisticated attackers who target critical systems using novel (not previously disclosed) tactics that training data do not reflect. This category includes conventional statistics-based and non-parametric methods. Behaviour-specification-based detection is a distinct, potentially most-effective variant for CPS, where humans manually define legitimate behaviour and the IDS flags deviations from it; it offers a low false-negative rate and immediate effectiveness (no training phase needed) and, like other behaviour-based methods, can catch zero-day attacks, but at the cost of the significant effort required to build the formal specification~\cite{mitchellSurveyIntrusionDetection2014}. However, such methods evaluate \gls{ids} without confidence score, detection-probability parameter, assigned per physical component from its task and goal criticality. Therefore, resilience evaluation and validation would be independent of any particular detection algorithm's internals.

Han et al.\ examine the particular constraints that \gls{cps-pl}
impose on detection resource limits, real-time requirements, and the interplay between network and physical anomalies,  noting that fault-tolerant mechanisms alone are insufficient when failures are deliberately induced rather than incidental~\cite{hanIntrusionDetectionCyberPhysical2014a}. 
Control-theoretic attack detection methods, such as those proposed by Pasqualetti et al., provide formal conditions to identify compromised actuators and sensors, but address identification rather than runtime response~\cite{pasqualettiAttackDetectionIdentification2013}. In the robotic domain specifically, as Zhu and his colleagues mentioned in their book, ``it is essential to see that OT-level safety and IT-level security are intertwined. The ignorance of IT-security will enable an attacker to take over the control of OT and create human-induced devastating incidents''. They provide a comprehensive treatment of cybersecurity challenges including vulnerability scoring, attack quantification, and the difficulty of hardening deployed robotic systems post-hoc, and identify runtime resilience as an open problem but do not provide a formal framework for reasoning about it at runtime~\cite{zhu2021roboticssecurity}.

Intrusion response work, such as the Autonomous Response Controller of Kholidy, addresses runtime response beyond detection but does not reason about disruption or degradation tolerability in terms of formal predicates~\cite{kholidyAutonomousMitigationCyber2021}.~\gls{robresilience} addresses this gap to determine whether a compromised system can continue operating and whether mitigation is feasible. 

\subsection[Fault tolerance and FDIR]{Fault tolerance and \gls{fdir}}
The engineering tradition of \gls{fdir} addresses operational continuity
under component failure~\cite{hwangSurveyFaultDetection2010}.
Model-based \gls{fdir} has been applied across safety-critical
domains including aerospace~\cite{zolghadriAdvancedModelbasedFDIR2012},
automotive drive-by-wire systems~\cite{isermannFaulttolerantDrivebywireSystems2002},
and power distribution networks~\cite{zidanFaultDetectionIsolation2017}.
However, \gls{fdir} is designed around deviations from expected physical behaviour and assumes that failure modes are known, bounded, and structurally representable. It does not model adversarial intent, criticality assignments that vary by task, or the compositional degradation effects that arise when multiple devices are compromised simultaneously through a coordinated cyberattack. 

Segovia-Ferreira et al.\ survey cyber-resilience approaches for \gls{cps-pl}, noting that research efforts have concentrated on intrusion detection, with little discussion of remediation once an intrusion is detected, and that most responses are manual or hardwired with fixed, non-configurable reactions~\cite{segovia-ferreiraSurveyCyberResilienceApproaches2024}. \gls{robresilience} addresses this by evaluating disruption and degradation tolerability as formally defined predicates over the compromised device set, instead of relying solely on detection output.

\subsection{Formal methods, cyber-resilience, and embodied systems}

Formal approaches to \gls{cps} security have produced frameworks
for modelling cyber-physical attacks using hybrid process
calculus~\cite{lanotteFormalApproachCyberPhysical2017} and for verifying obligation-preserving behaviour in autonomous agents~\cite{shea-blymyerAlgorithmicEthicsFormalization2021}. 

In the robotic systems domain, Kirca et al.\ propose a runtime verification architecture for \gls{ros2}-based security
monitoring, checking conformance to temporal logic specifications and detecting anomalies such as \gls{dos}-induced packet-count drops during
execution~\cite{kirca2023runtimeverification}. 
However, Kirca et al.\ address specification conformance and anomaly identification rather than formal predicate evaluation over a compromised device set with adaptive mitigation. Segovia-Ferreira et al.\ survey cyber-resilience approaches and identify the absence of runtime predicate validation frameworks as a recurring gap in the
literature~\cite{segovia-ferreiraSurveyCyberResilienceApproaches2024}. 

On the embodied
systems front, recent surveys document a rapidly expanding threat surface for autonomous platforms interacting with physical environments~\cite{xingRobustSecureEmbodied2026, perloEmbodiedAIEmerging2025}, but neither provides a formal predicate-based runtime
resilience mechanism with adaptive mitigation. Giaretta introduces exactly this: a formal framework defining tolerable disruption, tolerable degradation, and mitigation feasibility as runtime predicates for embodied \gls{cps-pl} under device-level
cyberattacks~\cite{giarettaFormalResilienceFramework2026}. \gls{robresilience} implements and evaluates this framework in a \gls{webots} with a \gls{pr2} robot under \gls{ros2}, providing the first empirical validation of its predicate semantics across all realisable disruption--degradation--mitigation
state combinations.

%% file: chapters/resilience_framework.tex
\section{Background: The Theoretical Resilience Framework}
\label{sec:framework}
As previously stated, this paper implements the formal resilience framework described in \cite{giarettaFormalResilienceFramework2026}. 
In the framework, a robotic system $R$ is composed of a finite set of devices $D = \{d_1, d_2, \ldots, d_n\}$. The robot can perform a finite set of tasks $T = \{t_1, t_2, \ldots, t_p\}$, and must preserve a finite set of embodiment-related goals $G = \{g_1, g_2, \ldots, g_q\}$. Each device $d \in D$ contributes to task execution and goal preservation through criticality mappings.

Following the same notation, $2^D$ denotes the power set of $D$, i.e., the set of all subsets of devices.
\color{black}

For the sake of brevity, in this section we report only the core theoretical concepts. The reader can find the full set of definitions and proofs in the original paper~\cite{giarettaFormalResilienceFramework2026}.
\color{black}

\begin{definition}
\label{def:tau}
The \textbf{task-criticality mapping} is defined as a function:
\[
  \tau : D \times T \to \mathcal{C},
\]
where $\mathcal{C} = \{0, 1, 2\}$ denotes the criticality levels: $0$ for none,
$1$ for important, and $2$ for required.
\end{definition}

\begin{definition}
\label{def:epsilon}
The \textbf{goal-criticality mapping} is also defined as a function:
\[
  \varepsilon : D \times G \to \mathcal{C},
\]
where $\mathcal{C} = \{0, 1, 2\}$ denotes the previously introduced criticality levels.
\end{definition}

\begin{definition}
\label{def:IDS}
Let $D$ be the set of devices present in the embodied \gls{cps}. \textbf{The \gls{ids} output} is a function
\[
  I : D \to [0, 1],
\]
where $I(d)$ denotes the confidence that the device $d \in D$ is currently compromised by an attack.
\end{definition}

\begin{definition}
\label{def:kappa}
Let $\kappa_{\mathrm{base}},\,\kappa_{\mathrm{crit}} \in [0,1]$ be fixed thresholds such that $\kappa_{\mathrm{crit}} < \kappa_{\mathrm{base}}$. The device-specific threshold function $\kappa : D \to [0, 1]$ is:
\[
  \kappa(d) =
  \begin{cases}
    \kappa_{\mathrm{crit}},
      & \text{if } \exists\,t \in T_{\mathrm{active}} : \tau(d,t) = 2 \\
      & \quad\text{or } \exists\,g \in G_{\mathrm{active}} : \varepsilon(d,g) = 2 \\[4pt]
    \kappa_{\mathrm{base}},
      & \text{otherwise,}
  \end{cases}
\]
where $T_{\mathrm{active}} \subseteq T$ and $G_{\mathrm{active}} \subseteq G$ are
the currently relevant tasks and embodiment-preserving goals, respectively.
\end{definition}

\begin{definition}
\label{def:S}
Given the \gls{ids} function $I(d)$ and the threshold function $\kappa(d)$, the current \textbf{set of compromised devices} $S$ is defined as:
\[
  S = \{\,d \in D \mid I(d) \geq \kappa(d)\,\}. \label{eq:compromised-set}
\]
\end{definition}

\begin{definition}
\label{def:delta}
The \textbf{tolerable disruption} is described as a function:
\[
  \delta : 2^{D} \to \{0, 1\},
\]
which evaluates whether a set of compromised devices $S \subseteq D$ includes any device that is critical (i.e., strictly required) for task execution or goal preservation. Let $T_{\mathrm{active}} \subseteq T$ and $G_{\mathrm{active}} \subseteq G$ be as defined in \Cref{def:kappa}. The predicate is defined as:
\[
  \begin{aligned}
    \delta(S) = 1 \;\iff\;
      &\Bigl(\forall\,t \in T_{\mathrm{active}},\;
             \nexists\,d \in S : \tau(d,t) = 2\Bigr) \\
      &\;\land\;
       \Bigl(\forall\,g \in G_{\mathrm{active}},\;
             \nexists\,d \in S : \varepsilon(d,g) = 2\Bigr).
  \end{aligned}
\]
This condition is satisfied only if no compromised device is critical for any active task or goal.
\end{definition}

\begin{definition}
\label{def:gamma}
Let 
\[
    \psi : 2^{D} \to [0,1]  
\]
be a monotonic non-increasing degradation function that evaluates system performance under a set of compromised devices $S \subseteq D$, where $1$ denotes full performance and $0$ denotes total failure. Let $\theta_{\mathrm{crit}},\,\theta_{\mathrm{base}} \in [0,1]$ be performance thresholds with $\theta_{\mathrm{crit}} > \theta_{\mathrm{base}}$.

The \textbf{tolerable degradation} is a function:
\[
  \gamma : 2^{D} \to \{0, 1\},
\]
which evaluates whether a system under disruption remains within acceptable performance bounds. The predicate is defined as:
\[
  \gamma(S) = 1 \;\iff\; \psi(S) \geq \theta(S),
\]
where the threshold $\theta(S)$ depends on the presence of critical devices:
\[
  \theta(S) =
  \begin{cases}
    \theta_{\mathrm{crit}},
      & \text{if } \exists\,d \in S,\;\exists\,t \in T_{\mathrm{active}} :
        \tau(d,t) = 2 \\
      & \quad\text{or } \exists\,d \in S,\;\exists\,g \in G_{\mathrm{active}} :
        \varepsilon(d,g) = 2 \\[4pt]
    \theta_{\mathrm{base}},
      & \text{otherwise.}
  \end{cases}
\]
The components of this condition are interpreted as follows:
\begin{itemize}
  \item $\gamma(S) = 1$: performance is within acceptable bounds;
  \item $\theta_{\mathrm{crit}}$: strict threshold applied when any critical device is affected;
  \item $\theta_{\mathrm{base}}$: relaxed threshold for non-critical device disruptions.
\end{itemize}
\end{definition}

\begin{definition}
Let $A$ be the set of symbolic mitigating actions, such as isolating compromised devices, reconfiguring system logic, or reassigning criticality to functionally equivalent non-compromised devices. Let $D_m \subseteq D$ denote the set of devices for which a mitigation action exists.

The \textbf{mitigation feasibility} is defined as a function:
\[
\label{def:mu}
  \mu : 2^{D} \to \{0, 1\},
\]
which evaluates whether a compromised set $S \subseteq D$ can be mitigated, at least partially, to restore sufficient system operability. Specifically:
\[
  \mu(S) = 1 \;\iff\; \label{eq:mu}
  \exists\,M \subseteq S \cap D_m :\;
  \delta(S \setminus M) = 1 \;\land\; \gamma(S \setminus M) = 1.
\]
This predicate returns true if there exists a subset of compromised devices $M$ for which an appropriate mitigation strategy can be applied such that both disruption and degradation become tolerable.
\end{definition}

Disruption ($\delta$) and degradation ($\gamma$) address two different aspects, and one does not imply the other. Disruption only looks at whether a critical device, required by the active task or goal, is compromised. Degradation looks at overall performance $\psi(S)$, which depends on every compromised device, critical or not. So several non-critical devices compromised at once can leave $\delta(S)=1$ true while still pushing $\psi(S)$ below $\theta_{\mathrm{base}}$, making $\gamma(S)=0$: the system can still attempt the task, but no longer performs it well enough. To be resilient without mitigations, a system needs both conditions to hold.
\color{black}

\begin{theorem}
Given a compromised set $S \subseteq D$, the \textbf{system is resilient} if and only if:
\[
\label{thm:resilience}
  \bigl(\delta(S) = 1 \;\land\; \gamma(S) = 1\bigr)
  \;\lor\;
  \mu(S) = 1.
\]
\end{theorem}

\begin{theorem}
\label{thm:resilience-posture}
Let $S$ be the current set of compromised devices as defined in \Cref{def:S}, derived from the \gls{ids} output $I(d)$ (\Cref{def:IDS}) and the threshold function $\kappa(d)$ (\Cref{def:kappa}). The \textbf{system's resilience posture} is evaluated according to the following cases:
\begin{enumerate}
  \item If $\delta(S) = 1$ and $\gamma(S) = 1$, then the system tolerates the
    disruption without degradation.
  \item Otherwise, if $\mu(S) = 1$, then the system is not disruption-tolerant,
    but a mitigation exists that restores tolerability, albeit with possible
    performance degradation.
  \item Otherwise, if $\mu(S) = 0$, then the disruption is not tolerable and the
    system must initiate safe-state mitigation or halt.
\end{enumerate}
\end{theorem}

For the proof sketches of ~\Cref{thm:resilience,thm:resilience-posture} we refer the reader to the theoretical framework paper on which this work is based on~\cite{giarettaFormalResilienceFramework2026}.
\color{black}

%% file: chapters/threat_model.tex
\section{Threat Model}
\label{sec:threat-model}
We consider an embodied CPS executing a task $t \in T$ 
autonomously, relying only on its navigation, manipulation, 
and perception subsystems to complete the task. No human 
operator is involved. In addition to these operational 
capabilities, the system is equipped with the resilience 
framework described in \Cref{sec:framework}, 
continuously evaluating $\delta$, $\gamma$, and $\mu$ 
during execution.

The adversary seeks to force $\delta(S) = 0$ or 
$\gamma(S) = 0$ by compromising one or more devices 
$d \in D$, violating the resilience guarantee of 
\Cref{thm:resilience}. Disruption may take the 
form of impeding navigation, interfering with object 
acquisition, or corrupting the robot's perception of the 
surrounding environment. The ultimate objective is to cause 
task failure, safety-relevant hazards such as collisions 
with nearby obstacles, or exhaustion of the available time 
budget.

The attacker operates at the \gls{ros2} communication layer, injecting symbolic message-based attacks targeting one or more devices $d \in D$ simultaneously. For each targeted 
device, an attack type is selected based on its overall objective. For example, if its objective is to tamper with obstacle avoidance, it will target devices involved in 
that subsystem, either individually or in a specific combination, to deviate from the original execution. Once an attack is active, it remains in force until a mitigation 
$M \subseteq S$ is deployed to resolve it, as formalised in \Cref{def:mu}.

The attacker is modelled as an outsider with partial system knowledge. It knows which devices $d \in D$ exist and which attack types are applicable to each. However, it has no 
access to the criticality mappings $\tau$ and $\varepsilon$, the internals of the Resilience Manager, or the \gls{ids} function $I$ defined in \Cref{sec:framework}. 
It cannot tamper with their execution or observe their outputs. The resilience framework and the \gls{ids} are assumed trusted and uncompromised, reflecting a deployment where both components run on isolated, hardened infrastructure.

\label{sec:defenses}
\paragraph{\textbf{Defenses:}} 
The \gls{ids} is probabilistic, providing a per-device confidence 
score $I(d) \in [0,1]$ that estimates the probability of an 
ongoing compromise.
This score reflects the cybersecurity status of each device. It captures only IDS-detectable cyberthreats, such as known-malware signatures, network reconnaissance (e.g., ARP scans), or denial-of-service flooding, rather than physical wear or non-malicious faults, which an IDS is not designed to recognise.
\color{black}
A device $d$ enters the compromised set 
$S$ when $I(d) \geq \kappa(d)$, as defined in \Cref{def:IDS,def:kappa,def:S}. The mitigations, when applied, are deterministic and fully 
resolve the ongoing attack for each targeted device, removing it from $S$ and restoring the evaluation of $\delta(S \setminus M)$ and $\gamma(S \setminus M)$.

%% file: chapters/implementation.tex
\section{Proposed Framework Implementation}
\label{sec:implementation}

The implementation translates the formal resilience framework 
presented in \Cref{sec:framework} into an executable 
evaluation system for embodied \gls{cps-pl}. The system operates 
continuously during task execution, reasoning about the current 
resilience state of the robot under cyberattack conditions 
consistent with the threat model of \Cref{sec:threat-model}. All the capital bold letters between round brackets in the following subsections, i.e. \textbf{(A)}, correspond to the connection labels in ~\Cref{fig:architecture}.

\subsection{System Architecture}
\label{sec:architecture}
\begin{figure*}[t] 
    \centering
    \includegraphics[width=\textwidth]{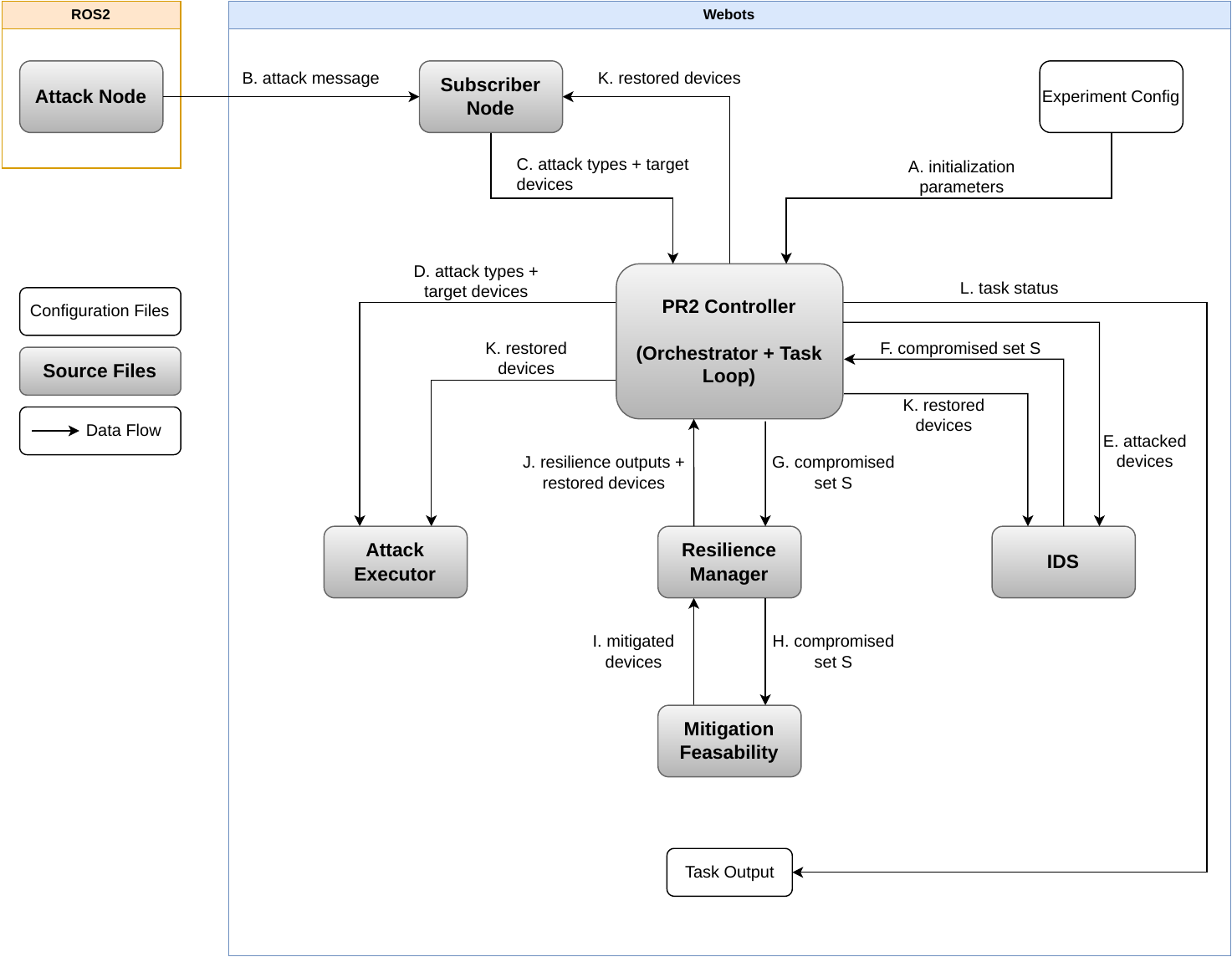}
    \caption{Overview of implementation and evaluation of the proposed framework. Source Files are Python modules, Data Flow depicts module's interaction, and Configuration Files presents different configuration setups. The diagram shows logical data flows between modules; the loop execution order and conditional invocation of each component
are described in ~\Cref{sec:implementation}.
    }
    \label{fig:architecture}
\end{figure*}

The framework is organized into four modules: a \textit{simulation
environment} hosting the physical robot, a \textit{coordination
controller} that orchestrates task execution and inter-module
communication, an \textit{attack injection module} responsible for
delivering and applying cyber-attacks, and a \textit{resilience
evaluation module} that continuously assesses the system state and
drives mitigation. \Cref{fig:architecture} illustrates the overall
architecture and the data flow at runtime.

During initialization, the controller loads the experiment 
configuration \textbf{(A)}, which specifies task type, goal activation, 
criticality mappings $\tau$ and $\varepsilon$, system thresholds 
$\kappa_{\mathrm{crit}}$, $\kappa_{\mathrm{base}}$, 
$\theta_{\mathrm{crit}}$, $\theta_{\mathrm{base}}$, reduction 
coefficients $\alpha_{\mathrm{crit}}$ and $\alpha_{\mathrm{base}}$, 
baseline execution time, and the mitigatable device set. These 
parameters are then distributed by the controller to the resilience evaluation module before task execution begins.


During runtime, the attack injection module introduces an attack by
publishing a symbolic message to the \texttt{active\_attacks} topic
\textbf{(B)}, received by the Subscriber Node and forwarded to the
coordination controller \textbf{(C)}. The controller dispatches the
attack information in two directions: to the Attack Executor
\textbf{(D)}, which immediately applies physical effects to the targeted
actuators, and to the IDS \textbf{(E)}, which performs probabilistic
detection. The IDS returns the resulting compromised set $S$ to the
controller \textbf{(F)}, which passes it to the Resilience Manager
\textbf{(G)}. The Resilience Manager evaluates $\delta$, $\gamma$,
and $\mu$, returning resilience outputs together with any restored
devices to the controller \textbf{(J)}. When mitigation is applied,
the controller broadcasts the restored device set simultaneously to
the Subscriber Node, the Attack Executor, and the IDS \textbf{(K)},
removing the affected devices from the active attack state across all modules. The final task result and runtime metrics are reported to the Task Output \textbf{(L)}.

The system is implemented in Python, using \gls{webots} as the simulation environment, a \gls{pr2} robot platform, and \gls{ros2} for attack communication. A key integration challenge arose from hardware contention: the original \gls{webots} \gls{pr2} controller 
continuously overwrites motor commands during task execution, silently negating attack effects. This was resolved by inserting attack-awareness checks into the low-level hardware control layer, granting the Attack Executor temporary ownership of targeted actuators when an attack is active.

\subsection{Framework Modules}
\label{sec:modules}
The four modules, described below, implement the core resilience predicates defined in \Cref{sec:framework}.


\subsubsection{Simulation Environment}
\label{sec:simulation}

The simulation environment is provided by \gls{webots}, which hosts
the \gls{pr2} robot and executes its physics-based dynamics. The PR2
is a full-scale service robot equipped with omnidirectional wheels,
two arms, grippers, and a head-mounted sensor suite. The Webots
Supervisor API grants the coordination controller programmatic access to the robot's actuators, sensors, and scene objects.

\subsubsection{Coordination Controller}
\label{sec:coordination}

The \gls{pr2} Controller acts as the central orchestrator of the
framework. It loads all modules from the Source Files and distributes
the experiment parameters \textbf{(A)}, manages modules'
communication at each control step, and integrates mitigation commands
with the robot's physical execution. Each control step is triggered by
the Task Loop invoking the resilience check during a motion primitive. At every step, it reads the
current attack state from the Subscriber Node \textbf{(C)}, drives
the Attack Executor \textbf{(D)} and the IDS \textbf{(E)}, receives
the compromised set $S$ \textbf{(F)}, forwards it to the Resilience
Manager \textbf{(G)}, and processes the resulting resilience outputs
\textbf{(J)}, including broadcasting any restored devices
\textbf{(K)}.


\paragraph{Task Loop}
\label{sec:task}

The Task Loop defines three robot behaviors: \textit{navigate to
goal}, \textit{pickup water bottle}, and the composite
\textit{navigate and pickup water bottle}, which combines grasping
with transport to a target table and controlled release. It is the
outermost control loop of the system: each task
is decomposed into smaller primitives that invoke the resilience check
at every step, propagating a \texttt{HALTED} signal immediately upon
failure so that execution is stopped as soon as possible.
A task that completes while the system is in a \texttt{NOT RESILIENT}
state is downgraded from \texttt{DONE} to \texttt{HALTED}, making sure
that only genuine recoveries are counted as successes. The task result
is reported to the Task Output \textbf{(L)}.

\subsubsection{Attack Injection Module}
\label{sec:attack-module}

The attack injection module works both with the \gls{ros2} communication layer
and the Webots simulation, and is responsible for introducing,
sustaining, and terminating cyber-attacks on the robot's components.
It comprises three components: the Attack Node, the Subscriber Node,
and the Attack Executor.

The \textbf{Attack Node} is a \gls{ros2} publisher operating outside
the Webots process, consistent with the external attacker model of
\Cref{sec:threat-model}. It publishes symbolic attack messages to the
\texttt{active\_attacks} topic \textbf{(B)}, specifying the target
device and attack type as a colon-separated string (e.g.,
\texttt{"left\_wheels:STOP"}).

The \textbf{Subscriber Node} runs inside the Webots process and
subscribes to the \texttt{active\_attacks} topic. Each received
message is parsed into a structured list of attack records and made
available to the coordination controller at the next step \textbf{(C)}.
When the controller broadcasts a restored device set \textbf{(K)},
the Subscriber Node removes the corresponding entries from its active
attack list, preventing mitigated devices from being injected again into the pipeline.


The \textbf{Attack Executor} receives attack types and target devices
from the controller \textbf{(D)} and applies their physical effects
to the corresponding Webots actuators. Attacks are typed symbolically:
\texttt{STOP} halts the targeted device entirely, \texttt{UNDERSPEED}
reduces actuator velocity, \texttt{OVERSPEED} increases actuator
velocity, \texttt{BACKWARD} reverses navigation direction, and
\texttt{GRIP\_WEAK} reduces gripper force. During an active attack,
the executor takes temporary hardware ownership of the targeted device,
preventing normal task commands from overwriting attack effects. When
the controller broadcasts the neutralized device set \textbf{(K)},
those devices are removed from the active attack list and their
hardware ownership is released.

\subsubsection{Resilience Evaluation Module}
\label{sec:resilience-module}

The resilience evaluation module runs entirely within the Webots
process and is responsible for detecting compromise, evaluating the
formal resilience predicates, and triggering mitigation. It comprises
two components: the IDS, which maintains the compromised set $S$, and
the Resilience Manager, which evaluates the framework predicates and
delegates feasibility search to its Mitigation Feasibility submodule.


\paragraph{IDS}
\label{sec:ids-module}

The \gls{ids} is evaluated at every control step. It receives the set of currently attacked devices from the
controller \textbf{(E)} and evaluates each device $d \in D$
independently against its device-specific detection threshold
$\kappa(d)$ as in \Cref{def:kappa}. Devices satisfying $\tau(d,t)=2$
or $\varepsilon(d,g)=2$ for any active task or goal are evaluated
against the stricter threshold $\kappa_{\mathrm{crit}}$; all others
use $\kappa_{\mathrm{base}}$. To mimic a real world implementation, detection rate (confidence) is modelled as a single trial per attack event: a uniform random draw is compared
against $\kappa(d)$, and a device enters $S$ only when the draw
exceeds the threshold. To prevent repeated evaluation of the same
ongoing attack from artificially inflating detection confidence, a
set with all the devices already tested ensures each device is assessed only once per
attack event, regardless of how many steps pass while the attack persists. The resulting compromised set $S$ is returned to the
controller \textbf{(F)}. When a device is restored \textbf{(K)},
it is removed from $S$ and enabled again for future detection.




\paragraph{Resilience Manager}
\label{sec:resilience-manager}

The Resilience Manager is evaluated at every control step. It receives the compromised set $S$ from the
controller \textbf{(G)} and evaluates $\delta$, $\gamma$, and $\mu$
as defined in \Cref{def:delta,def:gamma,def:mu}.

Four runtime states emerge from the combinations of $\delta$ and
$\gamma$. In the base state, $\delta(S)=1 \land \gamma(S)=1$: the
system is resilient and no mitigation is required, even with an
ongoing attack. When $\delta(S)=1 \land \gamma(S)=0$, no critical
device is compromised but aggregate degradation exceeds
$\theta_{\mathrm{base}}$; mitigations are deployed to restore
$\gamma$. When $\delta(S)=0 \land \gamma(S)=1$, a critical device
is compromised but $\psi(S)$ still exceeds $\theta_{\mathrm{crit}}$,
representing an edge case where the threshold configuration tolerates
the current degradation level. When $\delta(S)=0 \land \gamma(S)=0$,
both predicates are violated simultaneously, representing the most
severe state corresponding to a significant or coordinated attack.

In all states where $\delta(S)=0 \lor \gamma(S)=0$, the manager
delegates to the Mitigation Feasibility submodule. When $\mu(S)=1$,
a mitigation timer introduces a non-instantaneous delay before the
neutralized set $M \subseteq S$ is returned to the controller
\textbf{(J)} and broadcast to all active modules \textbf{(K)}. If no
feasible mitigation exists, the task returns \texttt{HALTED}.


\textbf{Mitigation Feasibility.}
Mitigation Feasibility evaluates whether the system can recover from
a disrupted state by applying mitigating actions to a subset of the
compromised devices. 
Unlike the IDS and Resilience Manager, which are evaluated at every
control step, Mitigation Feasibility is invoked conditionally: it
runs only when the Resilience Manager determines that the system is
not resilient and no mitigation is already pending.
The Resilience Manager passes the compromised
set $S$ \textbf{(H)}, and the submodule returns the neutralized
device subset \textbf{(I)}. The set of actionable devices is first computed as the
intersection of $S$ and the configured mitigatable device set. A
powerset is then enumerated, where each element represents a candidate
mitigation set $M$. For each candidate, a residual state
$S' = S \setminus M$ is constructed and evaluated using both the
disruption and degradation functions. If there exists a subset $M$
such that $\delta(S')=1$ and $\gamma(S')=1$, mitigation is considered
feasible: the submodule returns $\mu=1$ together with the neutralized
subset. If no such subset exists, it returns $\mu=0$.

\subsection{$\psi$ Function}
\label{sec:psi}
\begin{figure}[bt]
  \centering
  \includegraphics[width=0.98\columnwidth]%
    {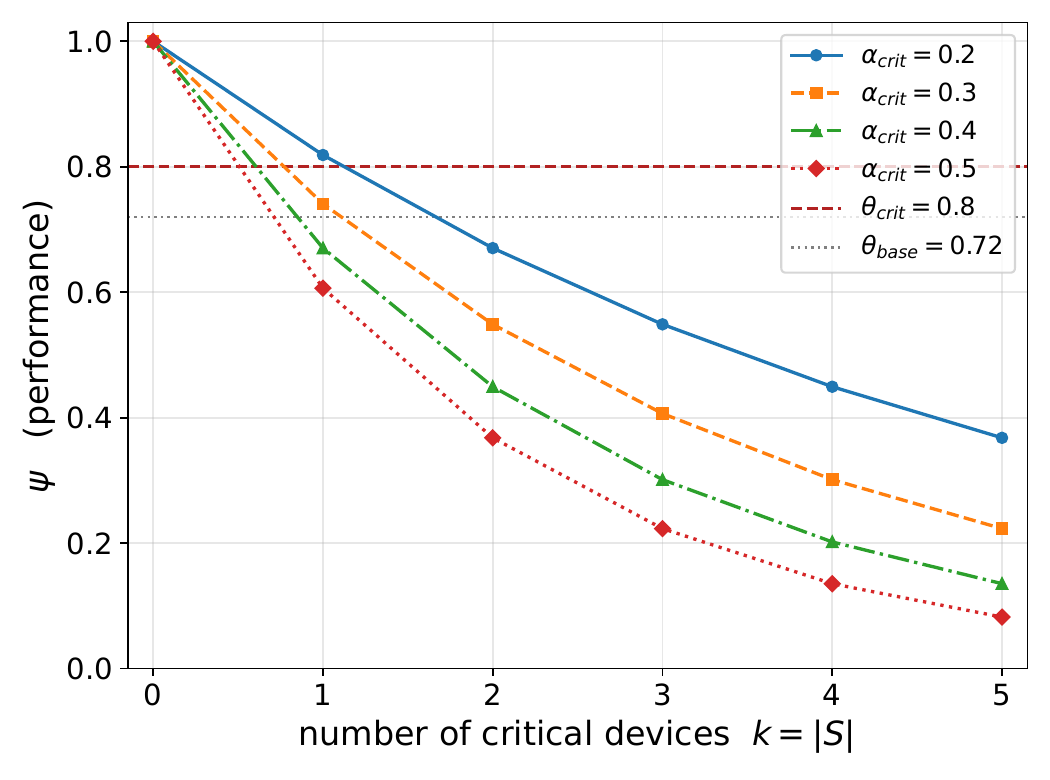}
  \caption{%
    Monotonic degradation of $\psi_{\mathrm{exp}}=e^{-\alpha_{crit} k_{crit}}$ as the
    number of compromised critical devices increases,
    shown for three values of $\alpha_{\mathrm{crit}}$.%
  }
  \label{fig:exp9_psi_monotonic}
\end{figure}

The degradation function $\psi : 2^{D} \rightarrow [0,1]$, 
introduced in \Cref{sec:framework}, is required to be 
monotonic and non-increasing but is not specified by the 
framework. The instantiation choice determines how aggressively 
compromised devices reduce the computed performance score, and 
therefore how quickly the system transitions into the degraded 
state $\gamma(S) = 0$. Let $k_{\mathrm{crit}} = |\{d \in S 
\mid \tau(d,t) = 2\}|$ and $k_{\mathrm{base}} = |\{d \in S 
\mid \tau(d,t) = 1\ \lor \tau(d,t) = 0\}|$ denote the number of compromised 
critical and non-critical devices respectively. The exponential form applies a steeper early penalty for 
critical device compromise as follows:
\begin{equation}
    \psi_{\mathrm{exp}}(S)
    = e^{-\left(
        \alpha_{\mathrm{crit}} \cdot k_{\mathrm{crit}}
        + \alpha_{\mathrm{base}} \cdot k_{\mathrm{base}}
    \right)}
    \label{eq:psi-exp}
\end{equation}
Because $D$ is finite, this instantiation produces a step function over the discrete set of device subsets regardless
of the underlying continuous form.
\Cref{fig:exp9_psi_monotonic} characterises the
general behaviour of the weighted exponential degradation
function $\psi_{\mathrm{exp}}$
(Equation~\eqref{eq:psi-exp}) across varying
$\alpha_{\mathrm{crit}}$ values.

The exponential form $\psi_{\mathrm{exp}}$ was selected for the experiments presented in \Cref{sec:result}, as its steeper initial slope better reflects the disproportionate operational impact of losing the first critical device. The specific parameter configurations used in the evaluation are reported in 
\Cref{sec:result}.

%% file: chapters/evaluation.tex
\section{Evaluation and Results}
\label{sec:result}

The evaluation is designed to verify the consistency between the
implementation and the theoretical framework of
\Cref{sec:framework}, and to characterise how the
configurable parameters introduced in
\Cref{sec:framework,sec:implementation}
influence runtime resilience behaviour under the attack
conditions defined in \Cref{sec:threat-model}.

\subsection{Experimental Setup}
\label{sec:experimental-setup}
All experiments are conducted in a \gls{webots} simulation
environment using a \gls{pr2} robot platform. The simulation scene comprises the robot, two wooden boxes serving as navigation
targets, one table containing a graspable object, and one
empty table used as a placement target. No physical hardware
is used; the evaluation is entirely simulation-based.

Each experiment is configured via a JSON file loaded by the
controller at initialisation, as described in
\Cref{sec:architecture}. The configuration specifies
task type, goal activation, baseline execution time,
criticality mappings $\tau$ and $\varepsilon$, system
thresholds $\kappa_{\mathrm{crit}}$ and
$\kappa_{\mathrm{base}}$, degradation thresholds
$\theta_{\mathrm{crit}}$ and $\theta_{\mathrm{base}}$,
reduction coefficients $\alpha_{\mathrm{crit}}$ and
$\alpha_{\mathrm{base}}$, and the mitigatable device set.
This configuration-driven design allows all parameters to be
varied between experiments without modifying the framework
code.

Attacks are automatically triggered during task execution by
publishing symbolic messages to the \gls{ros2}
\texttt{active\_attacks} topic, consistent with the attacker
capabilities of \Cref{sec:threat-model}.
Each attack message specifies one or more target devices
$d \in D$ and their associated attack type. The framework
logs all runtime state transitions, including updates to the
compromised set $S$, evaluations of $\delta$ and $\gamma$,
mitigation events, and task outcomes. Results are stored in
CSV files for post-hoc analysis. The collected metrics per
experiment are: task result (\texttt{DONE} or
\texttt{HALTED}), execution time, degradation percentage
relative to baseline, and the compromised set $S$ at each
evaluation step.

\subsection{Scenario Design and Coverage}
\label{sec:scenario-design}

The evaluation is structured around eight attack scenarios that together cover all realisable combinations of the three runtime predicates: tolerable disruption ($\delta$),
tolerable degradation ($\gamma$), and mitigation feasibility ($\mu$). Each scenario instantiates a distinct predicate configuration and is verified against the expected
theoretical behaviour defined in \Cref{sec:framework}. Scenarios are organised first by whether at least one critical device ($\tau = 2$) is compromised, determining whether $\delta = 0$ or $\delta = 1$ holds, then by degradation tolerability
($\gamma$), and finally by mitigation availability ($\mu$). \Cref{tab:scenario-overview} summarises all eight scenarios; 
\Cref{tab:experimental-setup} lists the parameters used in the experiments for each scenario; \color{black} \Cref{tab:predicate-coverage} maps the
$(\delta, \gamma)$ state space to the corresponding scenario identifiers, confirming that every cell of the predicate space is covered.

\begin{table*}[tb]
\caption{%
  Overview of the eight evaluation scenarios. Arrow notation ($0 \to 1$) denotes predicate recovery following successful mitigation. A dash in the $\mu$ column indicates that mitigation is not evaluated because the system never enters a
  non-resilient state. Parenthesised $\gamma$ values hold by construction in the
  time window shown but are not the primary displayed metric. Device abbreviations:
  LW\,=\,\texttt{left\_wheels}, RW\,=\,\texttt{right\_wheels}, LA\,=\,\texttt{left\_arm}, RA\,=\,\texttt{right\_arm}, LG\,=\,\texttt{left\_gripper}, RG\,=\,\texttt{right\_gripper}.
}
\label{tab:scenario-overview}
\renewcommand{\arraystretch}{1.2}
\setlength{\tabcolsep}{9pt}
\begin{tabular}{l l c c c c l r}
\toprule
\textbf{Scen.}
  & \textbf{Target devices}
  & \textbf{Critical}
  & $\boldsymbol{\delta}$
  & $\boldsymbol{\gamma}$
  & $\boldsymbol{\mu}$
  & \textbf{Outcome}
  & \textbf{Figure} \\
\midrule
1 & LA, RA & No  & $1$           & $1$          & ---
   & DONE           & \ref{fig:experiment-1_1} \\
2a & LA, RA & No  & $1$           & $0$          & $0$
   & HALTED          & \ref{fig:experiment-2a} \\
2b & LA, RA & No  & $1$           & $0 \to 1$    & $1$
   & DONE           & \ref{fig:experiment-2b} \\
3a & LW     & Yes & $0$           & $(1)$        & $0$
   & HALTED & \ref{fig:experiment-8}  \\
3b & LW, RW, LG, RA, LA, RG    & Yes & $0 \to 1$     & $(1)$        & $1$
   & HALTED & \ref{fig:experiment-7}  \\
3c & LW     & Yes & $0$           & $1$          & $0$
   & HALTED & \ref{fig:experiment-3c} \\
4a & LW, LA & Yes & $0$           & $0$          & $0$
   & HALTED           & \ref{fig:experiment-4a} \\
4b & LW, LA & Yes & $0 \to 1$          & $0 \to 1$    & $1$
   & DONE           & \ref{fig:experiment-4b} \\
\bottomrule
\end{tabular}
\end{table*}

\begin{table}[tb]
\caption{%
  Parameters in experimental setup for evaluation scenarios.
}
\label{tab:experimental-setup}
\renewcommand{\arraystretch}{1.2}
\setlength{\tabcolsep}{9pt}
\begin{tabular}{l c c c}
\toprule
\textbf{Scen.}
  & $\boldsymbol{\kappa_{crit}/ \kappa_{base}}$
  & $\boldsymbol{\theta_{crit}/ \theta_{base}}$
  & $\boldsymbol{\alpha_{crit}/ \alpha_{base}}$ \\
\midrule
1 & 0/0 & 0.8/0.72 & 0.15/0.2 \\
2a & 0/0 & 0.8/0.72 & 0.15/0.2 \\
2b & 0/0 & 0.8/0.72 & 0.15/0.2 \\
3a & 0/0 & 0.99/0.75 & 0.4/0.05 \\
3b & 0/0 & 0.99/0.75 & 0.4/0.05 \\
3c & 0/0 & 0.8/0.5 & 0.15/0.04 \\
4a & 0/0 & 0.8/0.72 & 0.15/0.2 \\
4b & 0/0 & 0.8/0.72 & 0.15/0.04 \\
\bottomrule
\end{tabular}
\end{table}

\color{black}

\begin{table}[tb] 
\caption{%
  Predicate coverage matrix.
  Each cell lists the scenarios realising the corresponding
  $(\delta, \gamma)$ combination;
  mitigation variants are distinguished by the $\mu$ value.%
}
\label{tab:predicate-coverage}
\large
\renewcommand{\arraystretch}{1.3}
\setlength{\tabcolsep}{8pt}
\begin{tabular}{c c c}
\toprule
& \multicolumn{1}{c}{$\boldsymbol{\gamma = 1}$}
& \multicolumn{1}{c}{$\boldsymbol{\gamma = 0}$} \\
\midrule
$\delta = 1$
  & 1
  & 2a $(\mu{=}0)$,\ 2b $(\mu{=}1)$ \\
$\delta = 0$
  & 3a $(\mu{=}0)$,\ 3b $(\mu{=}1)$,\ 3c
  & 4a $(\mu{=}0)$,\ 4b $(\mu{=}1)$ \\
\bottomrule
\end{tabular}
\end{table}

\subsection{Non-Disrupted Scenarios ($\delta = 1$)}
\label{sec:non-disrupted}

All scenarios in this group attack exclusively devices whose
task-level criticality satisfies $\tau \leq 1$, ensuring
that the disruption predicate $\delta$ remains $1$
throughout. The group isolates how degradation severity and
mitigation availability determine the overall resilience
classification when disruption is structurally precluded by
the criticality assignment.

\begin{figure*}[tb]
  \centering
  \begin{subfigure}[b]{0.48\textwidth}
    \centering
    \includegraphics[width=\columnwidth]%
      {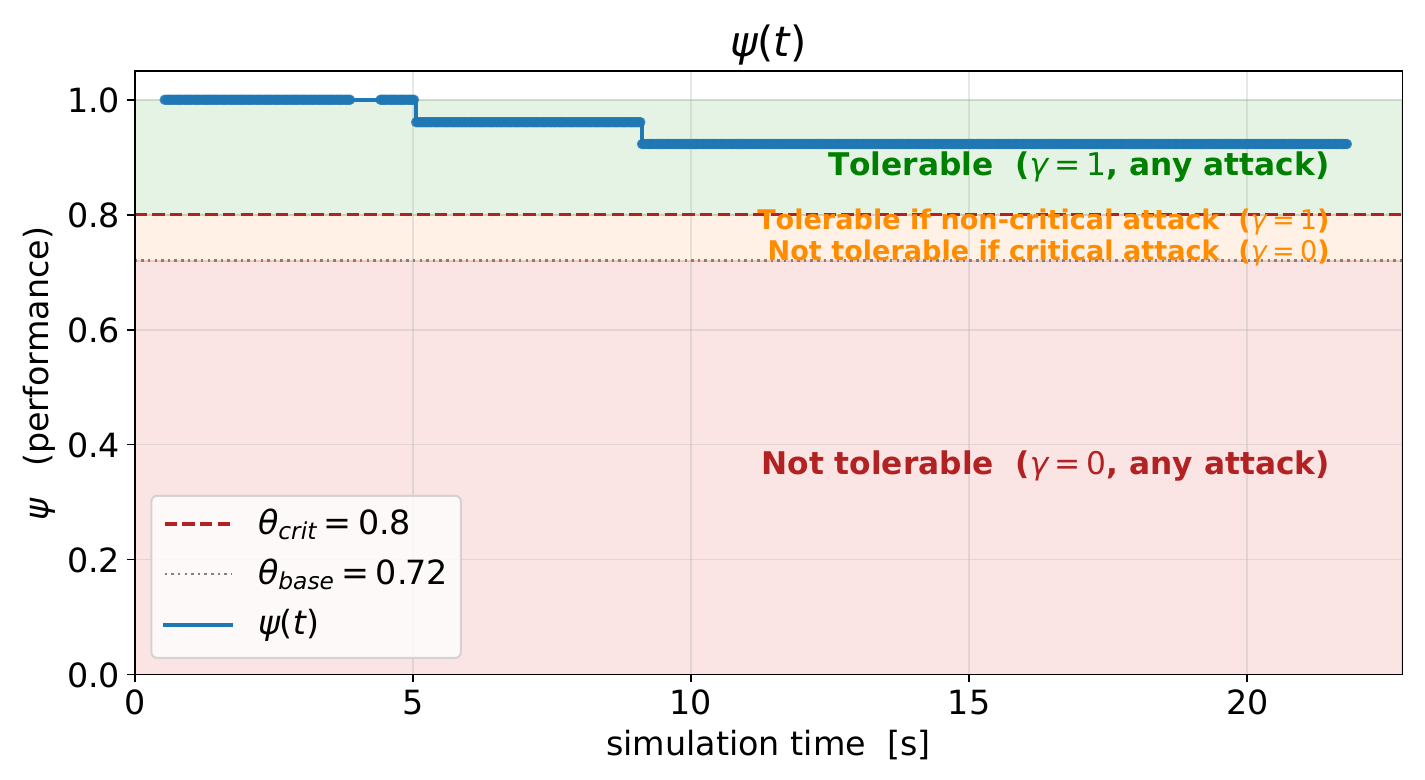}
    \caption{%
      Scenario~1: $\psi$ remains well within tolerance
      ($\gamma = 1$) under mild non-critical attack.%
    }
    \label{fig:experiment-1_1}
  \end{subfigure}%
  \hfill
  \begin{subfigure}[b]{0.48\textwidth}
    \centering
    \includegraphics[width=\columnwidth]%
      {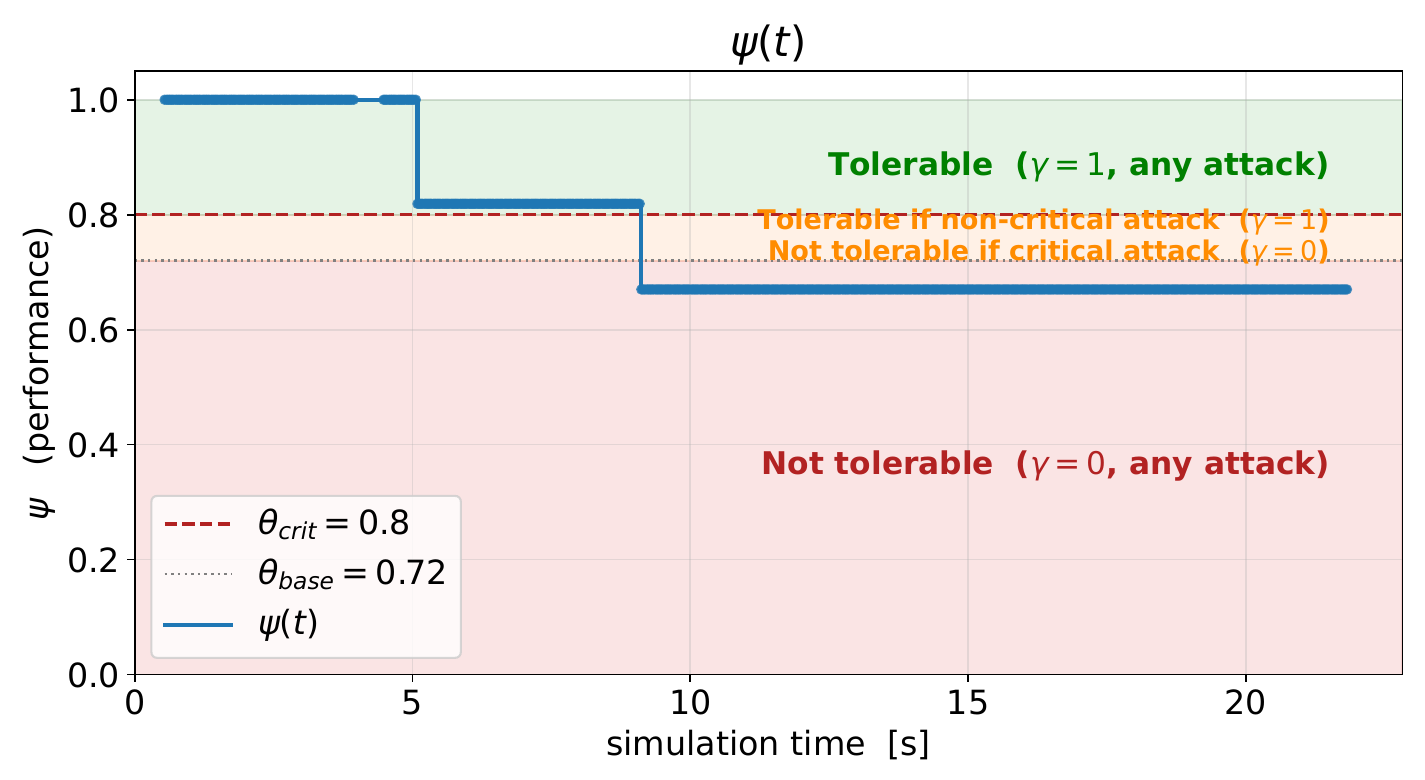}
    \caption{%
      Scenario~2a: higher degradation rate drives $\psi$
      below $\theta_{\mathrm{base}}$; $\gamma = 0$ persists
      with no mitigation available.%
    }
    \label{fig:experiment-2a}
  \end{subfigure}
  \caption{%
    Non-disrupted scenarios ($\delta = 1$) under
    non-critical device attacks (\texttt{left\_arm},
    \texttt{right\_arm}), without mitigation.
    Both runs share identical targets and thresholds;
    only $\alpha_{\mathrm{base}}$ differs.%
  }
  \label{fig:phi-non-crit}
\end{figure*}

\paragraph{Scenario 1: Resilient baseline.}
The \texttt{left\_arm} and \texttt{right\_arm} devices,
both non-critical ($\tau \leq 1$) for the active task (\textit{navigate to goal}), are
subjected to a \texttt{STOP} attack with a low base
degradation rate ($\alpha_{\mathrm{base}} = 0.04$,
$\theta_{\mathrm{crit}} = 0.8$, $\theta_{\mathrm{base}} =
0.72$, no mitigation). As shown in
\Cref{fig:experiment-1_1}, $\psi(t)$ remains at
approximately $0.93$ throughout the 160\,s run, well above
$\theta_{\mathrm{crit}}$. With both $\delta = 1$ and
$\gamma = 1$ satisfied at all times, the system remains in
the \emph{resilient} state and the task completes
successfully. This scenario serves as the baseline against
which the effect of increasing degradation severity and
enabling mitigation is measured in Scenarios~2a and~2b.

\paragraph{Scenario 2a: Intolerable degradation without mitigation.}
The same devices and attack type are repeated with a higher
base degradation rate ($\alpha_{\mathrm{base}} = 0.20$) and
no mitigation enabled ($\mu = 0$). As shown in
\Cref{fig:experiment-2a}, $\psi(t)$ crosses below
$\theta_{\mathrm{base}} = 0.72$ by $t \approx 9$\,s,
settling at approximately $0.67$, and the system transitions
to \emph{not tolerable} ($\gamma = 0$) for the remainder of
the run. \Cref{lst:non-crit-no-mit} reproduces the
corresponding Resilience Manager output, confirming that the
framework correctly identifies the intolerable state, reports
no available mitigation, and subsequently halts the task.

\begin{figure}[tb]
  \centering
  \begin{minipage}{\columnwidth}
    \begin{lstlisting}[
      style=terminal,
      caption={%
        Scenario~2a: non-critical attack without mitigation.
        The second attack wave drives $\psi$ below
        $\theta$ and the task is halted.%
      },
      label={lst:non-crit-no-mit},
      basicstyle=\small\ttfamily,
      breaklines=true,
      literate=
        {ψ}{{$\psi$}}1
        {θ}{{$\theta$}}1
        {δ}{{$\delta$}}1
        {γ}{{$\gamma$}}1
        {→}{{$\rightarrow$}}1
        {<}{{$<$}}1
        {>}{{$>$}}1
        {=}{{$=$}}1]
[RM] System not disrupted → δ = 1
[RM] Degradation within tolerance → γ = 1
[RM] Resilience state: RESILIENT
ATTACK: left_arm (STOP) - NON-CRITICAL
[RM] Compromised set S updated: {'left_arm'}
ATTACK: right_arm (STOP) - NON-CRITICAL
[RM] Compromised set S updated: {'left_arm', 'right_arm'}
[RM] System degraded beyond tolerance: ψ=0.67 < θ=0.72 → γ = 0
[RM] Resilience state: NOT RESILIENT
[RM] No mitigation possible - system irreparably compromised
Task result: HALTED
    \end{lstlisting}
  \end{minipage}
\end{figure}

\begin{figure}[tb]
  \centering
  \includegraphics[width=\columnwidth]%
    {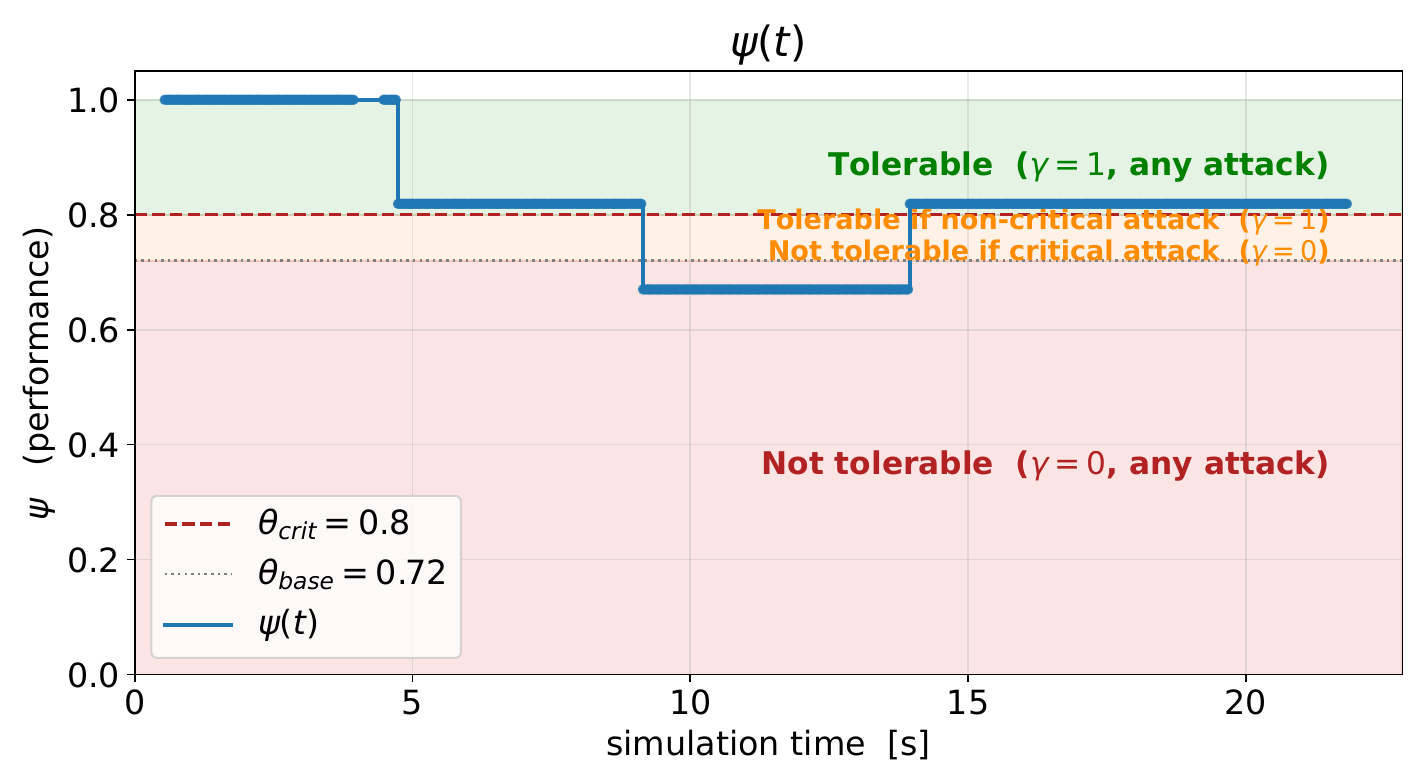}
  \caption{%
    Scenario~2b: mitigation arrests the initial $\psi$ drop
    caused by non-critical device attacks and restores
    $\gamma = 1$.%
  }
  \label{fig:experiment-2b}
\end{figure}

\paragraph{Scenario 2b: Mitigation restores tolerability.}
This scenario repeats the configuration of Scenario~2a with
mitigation enabled for the attacked devices. As shown in
\Cref{fig:experiment-2b}, $\psi(t)$ undergoes the same
initial drop to approximately $0.67$, briefly crossing below
$\theta_{\mathrm{base}}$. The mitigation loop then neutralises
\texttt{left\_arm}, removes it from the compromised set $S$,
and recovers $\psi(t)$ to approximately $0.81$, above
$\theta_{\mathrm{base}} = 0.72$, where it stabilises for the
remainder of the run ($\gamma = 1$). \Cref{lst:non-crit-mit}
confirms this recovery sequence in the Resilience Manager
output.

\begin{figure}[b]
  \centering
  \begin{minipage}{\columnwidth}
    \begin{lstlisting}[
      style=terminal,
      caption={%
        Scenario~2b: non-critical attack with mitigation.
        The Resilience Manager neutralises \texttt{left\_arm} and restores $\gamma = 1$.%
      },
      label={lst:non-crit-mit},
      basicstyle=\small\ttfamily,
      breaklines=true,
      literate=
        {ψ}{{$\psi$}}1
        {θ}{{$\theta$}}1
        {δ}{{$\delta$}}1
        {γ}{{$\gamma$}}1
        {→}{{$\rightarrow$}}1
        {<}{{$<$}}1
        {>}{{$>$}}1
        {=}{{$=$}}1]
[RM] System not disrupted → δ = 1
[RM] Degradation within tolerance → γ = 1
[RM] Resilience state: RESILIENT
ATTACK: left_arm (STOP) - NON-CRITICAL
[RM] Compromised set S updated: {'left_arm'}
ATTACK: right_arm (STOP) - NON-CRITICAL
[RM] Compromised set S updated: {'left_arm', 'right_arm'}
[RM] System degraded beyond tolerance: ψ=0.67 < θ=0.72 → γ = 0
[RM] Resilience state: NOT RESILIENT
[RM] Mitigation pending
[RM] Active mitigation neutralizing: {'left_arm'}
[RM] Compromised set S updated: {'right_arm'}
[RM] Degradation within tolerance → γ = 1
[RM] Resilience state: RESILIENT
Task result: DONE
    \end{lstlisting}
  \end{minipage}
\end{figure}

Scenarios~1--2b together confirm that the criticality
gating embedded in $\delta$ operates correctly: no attack
confined to non-critical devices triggered the disruption
path in any tested configuration. The paired runs of
Scenarios~2a and~2b further show that, in the
non-disrupted regime, the degradation and mitigation logic
resolve independently of $\delta$ as specified by the
framework.

These results also show that even a single non-critical device, 
when added to an already-degraded system, 
can tip $\psi$ across the baseline threshold if the
degradation rate is sufficiently high.
The resilience is sensitive to the
cumulative composition of the compromised set $S$, not only
to the criticality of individual devices.

\subsection{Disrupted Scenarios ($\delta = 0$)}
\label{sec:disrupted}
The remaining scenarios each compromise at least one device
with $\tau = 2$, guaranteeing $\delta = 0$ upon attack
regardless of the current degradation level. These scenarios
exercise the framework's behaviour under confirmed disruption,
including the secondary evaluation of $\gamma$ and the effect
of mitigation on predicate recovery.

\subsubsection{Tolerable Degradation ($\gamma = 1$)}
\label{sec:tolerable-degradation}

Scenarios~3a, 3b, and 3c each produce $\delta = 0$ without
$\gamma$ reaching $0$. They demonstrate that disruption and
non-resilience are not synonymous in the framework: a system
can remain within acceptable degradation bounds even after
losing a critical component, and mitigation can restore
$\delta$ without $\gamma$ having been violated.

\begin{figure}[tb]
  \centering
  \begin{subfigure}[b]{\columnwidth}
    \centering
    \includegraphics[width=\columnwidth]%
      {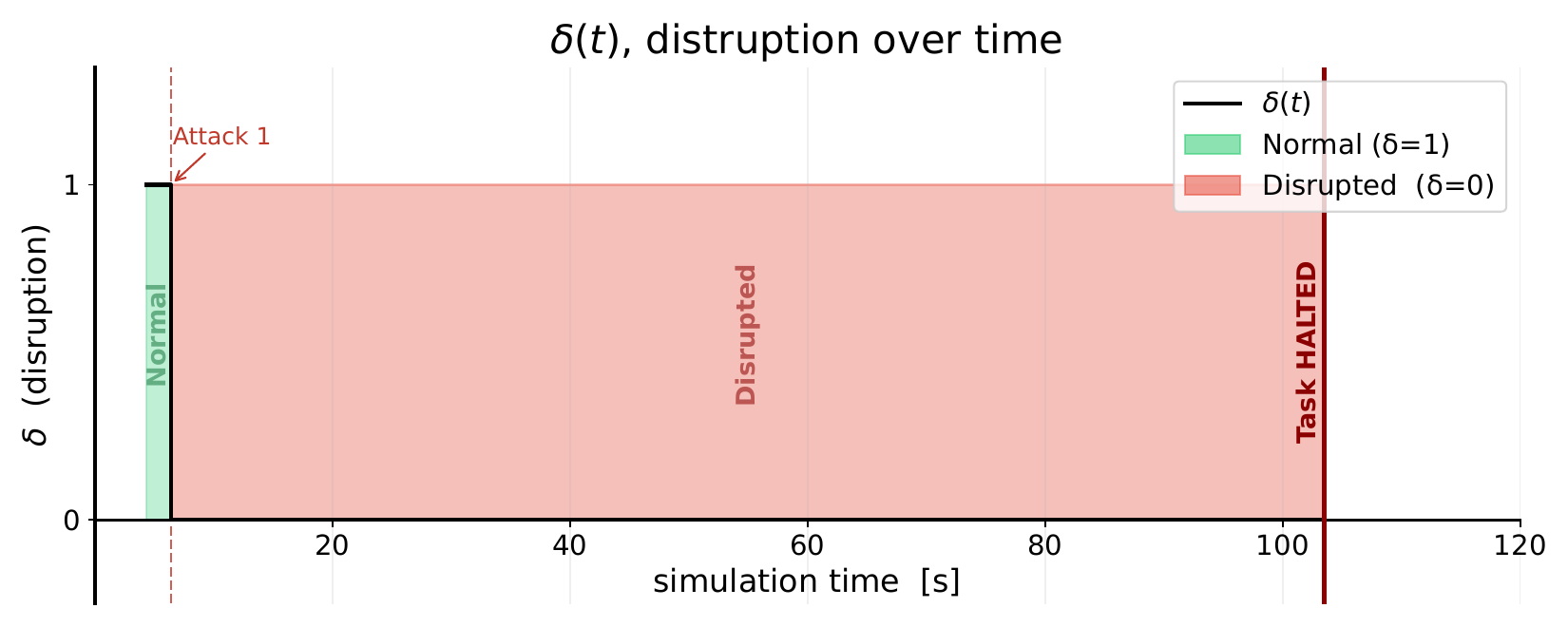}
    \caption{%
      Scenario~3a: $\delta$ over time, critical device
      (\texttt{left\_wheels}), no mitigation.
      $\delta = 0$ persists until task halt.%
    }
    \label{fig:experiment-8}
  \end{subfigure}
  \par\bigskip
  \begin{subfigure}[b]{\columnwidth}
    \centering
    \includegraphics[width=\columnwidth]%
      {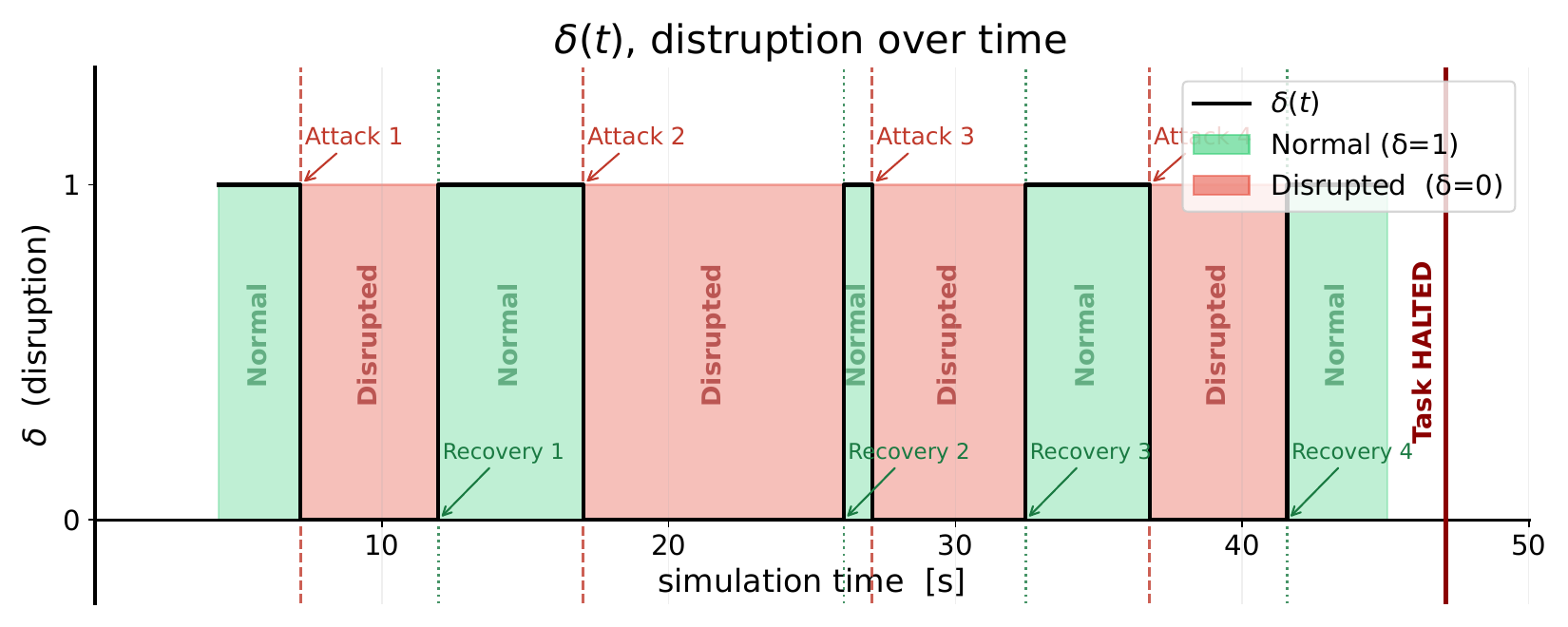}
    \caption{%
      Scenario~3b: same configuration with full mitigation.
      Repeated attack--recovery cycles drive $\delta$
      back to $1$.%
    }
    \label{fig:experiment-7}
  \end{subfigure}
  \caption{%
    Comparison of $\delta$ over time for a critical-device
    attack: (a)~without mitigation, $\delta$ remains
    permanently at $0$; (b)~with mitigation, the framework
    restores $\delta = 1$ across repeated attack--recovery
    cycles.%
  }
  \label{fig:delta_comparison}
\end{figure}

\paragraph{Scenarios 3a and 3b: Disruption with and without mitigation.}
The \texttt{left\_wheels}, defined as critical for task \textit{navigate and pick up water bottle}, are subjected to a \texttt{STOP} attack.
\Cref{fig:experiment-8,fig:experiment-7}
share identical task and parameter configurations but differ
in mitigation scope. In Scenario~3a
(\Cref{fig:experiment-8}), mitigation is unavailable;
a single attack at $t \approx 5$\,s leaves the robot
permanently disrupted ($\delta = 0$), no recovery event
occurs and the task runs until
\texttt{HALTED} at $t \approx 103$\,s. In Scenario~3b
(\Cref{fig:experiment-7}), full mitigation coverage is
enabled; four attack--recovery cycles occur within the
48.8\,s baseline, the task is restored and the state is \texttt{DONE}. In both
figures, this distinction has a decisive effect on the
$\delta$ trajectory: with mitigation, $\delta$ is repeatedly
driven back to $1$ within each cycle; without it, $\delta$
remains permanently at $0$.

\begin{figure}[tb]
  \centering
  \includegraphics[width=\columnwidth]%
    {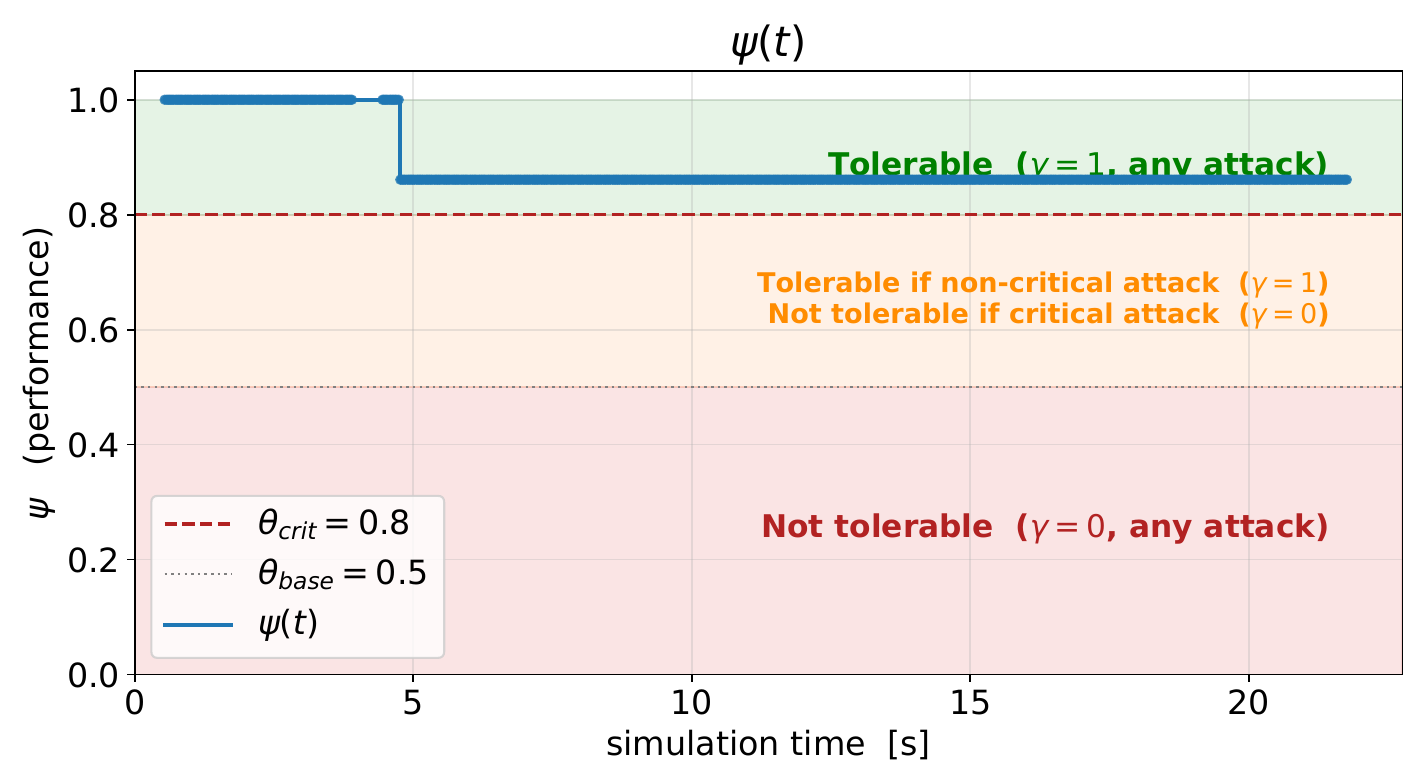}
  \caption{%
    Scenario~3c: $\psi$ remains above $\theta_{\mathrm{crit}}$
      throughout ($\gamma = 1$, $\delta = 0$). Mild
      degradation rate keeps the system tolerable despite
      disruption.%
  }
  \label{fig:experiment-3c}
\end{figure}

\paragraph{Scenario 3c: Degradation remains tolerable under disruption.}
\Cref{fig:experiment-3c} examines a critical-device
attack (\texttt{left\_wheels}, $\tau = 2$, \textit{navigate to goal}) from the $\psi$
perspective, with $\alpha_{\mathrm{crit}} = 0.15$,
$\alpha_{\mathrm{base}} = 0.04$, $\theta_{\mathrm{crit}} =
0.8$, and no mitigation. Following the attack, $\psi$ sustains
a single-step drop to approximately $0.86$, remaining above
$\theta_{\mathrm{crit}}$ throughout the run and yielding a
tolerable outcome ($\gamma = 1$). This scenario establishes
that $\delta = 0$ does not imply $\gamma = 0$: when
degradation is mild enough to remain within the critical
threshold, the system could technically continue operating in a disrupted
but tolerable degraded state. However, due to the resilience requirements defined
in \Cref{sec:framework} with \Cref{thm:resilience}, $\delta = 0$ must
 be restored to 1, but no mitigation is available. For this reason, the task status is \texttt{HALTED}.

\subsubsection{Intolerable Degradation ($\gamma = 0$)}
\label{sec:intolerable-degradation}

Scenarios~4a and 4b represent the most severe predicate
combination: $\delta = 0$ and $\gamma = 0$ simultaneously.
They verify that the framework correctly identifies
irrecoverable failure states and that mitigation can
resolve them when available.

\begin{figure}[tb]
  \centering
  \includegraphics[width=\columnwidth]%
    {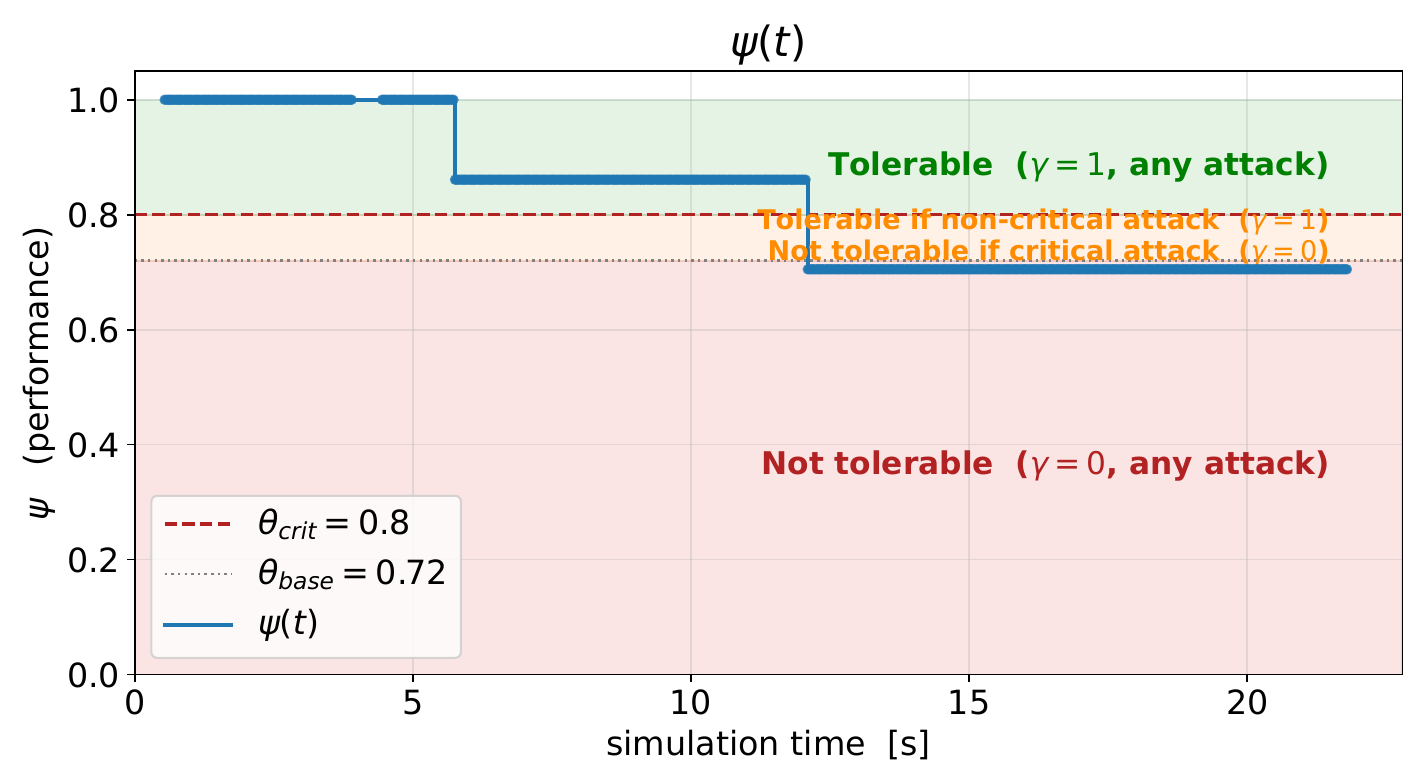}
  \caption{%
    Scenario~4a: escalating wheel and arm devices attack drives $\psi$
      below $\theta_{\mathrm{crit}}$ ($\gamma = 0$,
      $\delta = 0$). No mitigation is available.%
  }
  \label{fig:experiment-4a}
\end{figure}

\paragraph{Scenario 4a: Permanent non-resilience without mitigation.}
\Cref{fig:experiment-4a} applies a two-wave attack on
the \texttt{navigate\_to\_goal} task with a high degradation
rate ($\alpha_{\mathrm{crit}} = 0.15$,
$\alpha_{\mathrm{base}} = 0.20$, $\theta_{\mathrm{crit}} =
0.8$) and no mitigation.

The first wave (\texttt{left\_wheels}, critical) drives $\psi$ to approximately
$0.9$, just above $\theta_{\mathrm{crit}}$; the second wave (\texttt{left\_arm}, non-critical) pushes $\psi$ further to approximately $0.7$, below $\theta_{\mathrm{crit}}$,
placing the system unconditionally in the critical-device-not-tolerable region ($\gamma = 0$). Even if the second device attacked is not critical, the threshold used is $\theta_{\mathrm{crit}}$ due to the presence of a critical device. \Cref{lst:crit-no-mit} reproduces the Resilience Manager output, confirming that both disruption and intolerable degradation are identified \texttt{HALTED} and the task is stopped 27.0\,s before the baseline.

\begin{figure}[tb]
  \centering
  \begin{minipage}{\columnwidth}
    \begin{lstlisting}[
      style=terminal,
      caption={%
        Scenario~4a: critical-device attack without
        mitigation. Escalating wheel compromise drives
        both $\delta = 0$ and $\gamma = 0$.%
      },
      label={lst:crit-no-mit},
      basicstyle=\small\ttfamily,
      breaklines=true,
      literate=
        {ψ}{{$\psi$}}1
        {θ}{{$\theta$}}1
        {δ}{{$\delta$}}1
        {γ}{{$\gamma$}}1
        {→}{{$\rightarrow$}}1
        {<}{{$<$}}1
        {>}{{$>$}}1
        {=}{{$=$}}1]
[RM] System not disrupted → δ = 1
[RM] Degradation within tolerance → γ = 1
[RM] Resilience state: RESILIENT
ATTACK: left_wheels (UNDERSPEED) - CRITICAL
[RM] Compromised set S updated: {'left_wheels'}
[RM] System disrupted → δ = 0
[RM] Resilience state: NOT RESILIENT
[RM] No mitigation possible - system irreparably compromised
ATTACK: left_arm (UNDERSPEED) - NON-CRITICAL
[RM] Compromised set S updated: {'left_arm', 'left_wheels'}
[RM] System degraded beyond tolerance: ψ=0.7 < θ=0.8 → γ = 0
Task result: HALTED
    \end{lstlisting}
  \end{minipage}
\end{figure}

\begin{figure}[tb]
  \centering
  \includegraphics[width=\columnwidth]%
    {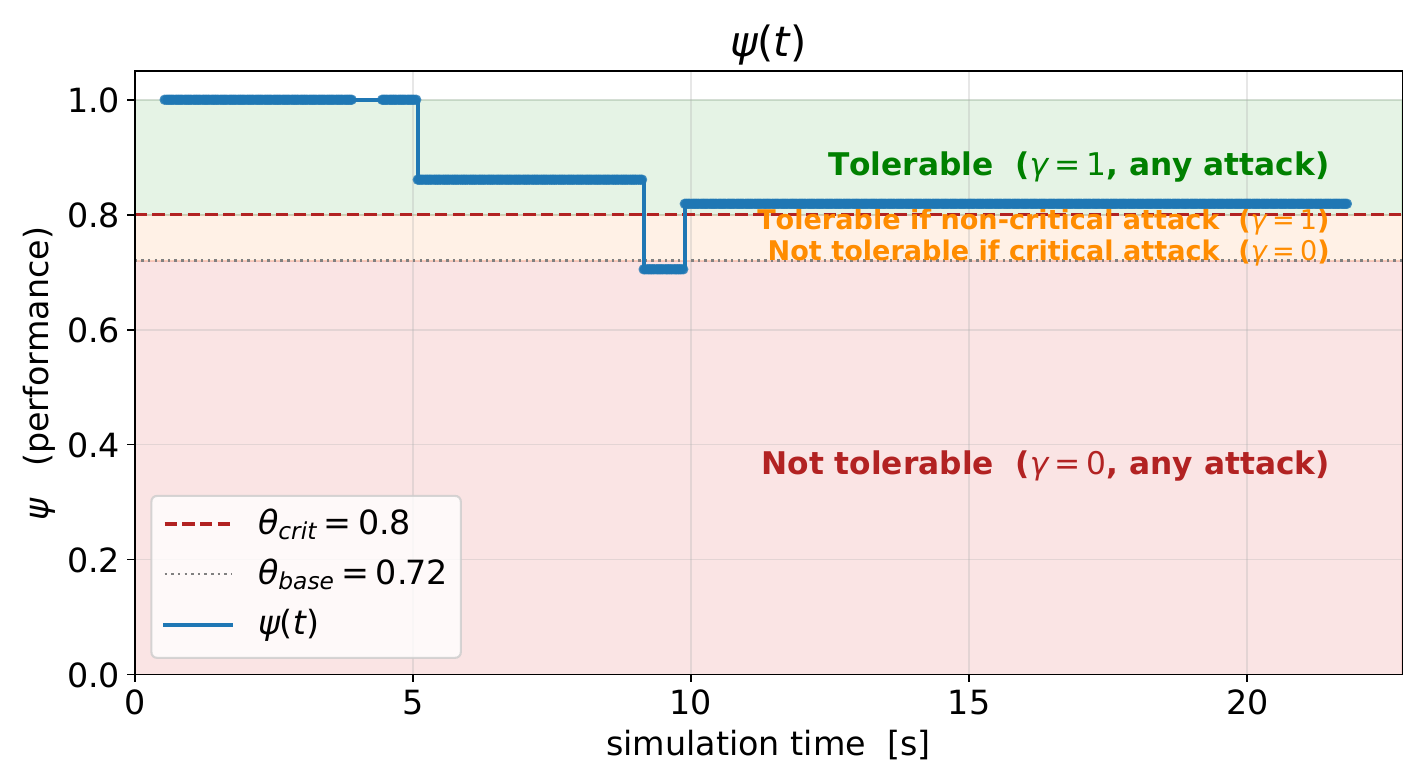}
  \caption{%
    Scenario~4b: $\psi$ timeline for critical and non-critical devices
    attack with mitigation. The initial drop below
    $\theta_{\mathrm{crit}}$ is arrested by the mitigation
    loop, which restores $\delta = 1$ and $\gamma = 1$.%
  }
  \label{fig:experiment-4b}
\end{figure}

\paragraph{Scenario 4b: Mitigation under simultaneous predicate failure.}
\Cref{fig:experiment-4b} applies the same attack
pattern as Scenario~4a with mitigation enabled for left wheel. The first part of the attack drives $\psi$ below $\theta_{\mathrm{crit}}$ (attack both critical \texttt{left\_wheels} and non-critical \texttt{left\_arm}), triggering $\gamma = 0$ alongside $\delta = 0$. The mitigation loop then neutralises the compromised device \texttt{left\_wheels} and removes it from $S$; $\psi$ returns above $\theta_{\mathrm{crit}}$, restoring both $\delta = 1$ and $\gamma = 1$ simultaneously.

\Cref{lst:crit-mit} confirms this recovery sequence:
the Resilience Manager transitions from \texttt{NOT RESILIENT} back to \texttt{RESILIENT} and the task completes
with a \texttt{DONE} outcome.

\begin{figure}[tb]
  \centering
  \begin{minipage}{\columnwidth}
    \begin{lstlisting}[
      style=terminal,
      caption={%
        Scenario~4b: critical-device attack with mitigation.
        The Resilience Manager restores $\delta = 1$ and
        $\gamma = 1$ following successful device
        neutralisation.%
      },
      label={lst:crit-mit},
      basicstyle=\small\ttfamily,
      breaklines=true,
      literate=
        {ψ}{{$\psi$}}1
        {θ}{{$\theta$}}1
        {δ}{{$\delta$}}1
        {γ}{{$\gamma$}}1
        {→}{{$\rightarrow$}}1
        {<}{{$<$}}1
        {>}{{$>$}}1
        {=}{{$=$}}1]
[RM] System not disrupted → δ = 1
[RM] Degradation within tolerance → γ = 1
[RM] Resilience state: RESILIENT
ATTACK: left_wheels (UNDERSPEED) - CRITICAL
[RM] Compromised set S updated: {'left_wheels'}
[RM] System disrupted → δ = 0
[RM] Resilience state: NOT RESILIENT
[RM] Mitigation pending
ATTACK: left_arm (UNDERSPEED) - NON-CRITICAL
[RM] Compromised set S updated: {'left_arm', 'left_wheels'}
[RM] System degraded beyond tolerance: ψ=0.7 < θ=0.8 → γ = 0
[RM] Active mitigation neutralizing: {'left_wheels'}
[RM] Compromised set S updated: {'left_arm'}
[RM] System not disrupted → δ = 1
[RM] Degradation within tolerance → γ = 1
[RM] Resilience state: RESILIENT
Task result: DONE
Task execution time increased with 1.8 seconds
Degradation: 9.0%
    \end{lstlisting}
  \end{minipage}
\end{figure}

\subsection{Limitations}
\label{sec:experimental-limitations}
To isolate the resilience logic from \gls{ids} uncertainty, we set both
$\kappa_{\mathrm{crit}}$ and $\kappa_{\mathrm{base}}$ to $0$ across all scenarios.
This choice deliberately contradicts
\Cref{def:kappa} in \Cref{sec:framework}, which
requires $\kappa_{\mathrm{crit}} < \kappa_{\mathrm{base}}$:
setting both parameters to $0$ violates this strict inequality.
We made this choice to decouple the evaluation of $\delta$,
$\gamma$, and $\mu$ from detection variability. The practical
implication is that any attacked device is guaranteed to enter
the compromised set $S$. While this is appropriate for
correctness verification, it undermines two important aspects of
the framework.
The first is logical. The strict inequality
$\kappa_{\mathrm{crit}} < \kappa_{\mathrm{base}}$ encodes the
requirement that the system declare a critical device compromised
with less evidence than it would require for a base device. If we
relaxed the inequality to
$\kappa_{\mathrm{crit}} \leq \kappa_{\mathrm{base}}$, critical and
base devices could share the same confidence threshold, defeating
the purpose of the criticality mapping. The second is practical:
in realistic deployment conditions, the \gls{ids} operates with
imperfect detection rates, so assuming 100\% accuracy is
unrealistic.

%% file: chapters/future_work.tex
\section{Future Work}
\label{sec:futurework}
The most immediate extension of this work would be to determine how to obtain realistic values for the thresholds and criticality mappings, rather than pre-determine them. One possibility is to learn the mappings via machine learning, calibrating parameters in simulation and then transferring them to the embodied CPS via a sim-to-real process. For example, a model could provide the values based on the type of device, its specifications, and the defined tasks and goals.
\color{black}

The evaluation also rests on several simplifications that we leave
to future work. We inject attacks symbolically rather than as real
\gls{ros2} exploits, bypass the IDS entirely by setting $\kappa = 0$, and model mitigations as deterministic. Of these, the IDS bypass is the one we view as most pressing to address. 

Future work should investigate the effect of varying $\kappa_{\mathrm{crit}}$ and $\kappa_{\mathrm{base}}$ on the runtime evaluations of $\delta$, $\gamma$, and $\mu$, replacing the perfect-detection assumption with a genuine probabilistic IDS. We recommend conducting multiple runs per scenario and reporting the mean and standard deviation of $\psi(t)$ and the resilience state transitions over time. A reasonable hypothesis is that outcome variance increases as detection rate decreases: lower accuracy produces more variable compositions of $S$, leading to less predictable predicate evaluations and resilience classifications. 

Replacing the remaining two simplifications, symbolic attack injection and deterministic mitigations, with real middleware-level attacks and realistic mitigation mechanisms would allow the full stochastic behaviour of the predicate loop to be studied across multiple runs and robot platforms.

Finally, we have only tested the framework in simulation. Physical
deployment on a real robot would expose timing and noise
constraints that \gls{webots} does not capture, and would determine
whether the Resilience Manager can meet the real-time constraints imposed by  embodied CPSs.
\color{black}

%% file: chapters/conclusion.tex
\balance

\section{Conclusion}
\label{sec:conclusion}
The results across all eight scenarios confirm that the
implemented framework correctly instantiates the theoretical
definitions of \Cref{sec:framework} at runtime.
Three structural properties of the implementation can be
verified from the experimental record.

First, the criticality gating embedded in $\delta$ operates
as specified. No attack confined to non-critical devices
($\tau \leq 1$) triggered the disruption path in any
scenario: Scenarios~1, 2a, and 2b each maintained
$\delta = 1$ regardless of attack severity or mitigation
status. Conversely, every attack targeting a device with
$\tau = 2$ immediately drove $\delta = 0$, confirming that
the compromised-set formation
(\Cref{def:S}) and the disruption
check (\Cref{def:delta}) are consistent with the
formal definition.

Second, mitigation correctly restores resilience across all
predicate combinations in which it was evaluated. In
Scenario~2b ($\delta = 1$, $\gamma: 0 \to 1$), mitigation
acted solely on the degradation predicate, neutralising the
non-critical compromise and recovering $\gamma$ while
$\delta$ was unaffected. In Scenario~3b
($\delta: 0 \to 1$, $\gamma = 1$), mitigation restored
disruption tolerance without $\gamma$ having been violated.
In Scenario~4b ($\delta: 0 \to 0$, $\gamma: 0 \to 1$), mitigation
resolved both predicates simultaneously, producing the only
case in which a doubly-failed system was fully recovered.
Taken together, these three cases establish that the
mitigation loop correctly traverses the predicate space as
specified in \Cref{def:mu}.

Third, the $\delta$ and $\gamma$ predicates are evaluated
independently, and their combination determines the resilience
classification without conflation. Scenario~3c illustrates
this most directly: with $\delta = 0$ and $\gamma = 1$
simultaneously, the system operated in a disrupted but
tolerable degraded state for the duration of its time budget. This is
the operational distinction between \emph{degraded} and
\emph{non-resilient} that the framework preserves: losing a
critical component does not automatically imply that
performance has become unacceptable. However, per the formal definition of resilience, an unrestored critical device still triggers a \texttt{HALTED} system state, preventing further damage.

Finally, the cumulative composition of the compromised set
$S$ is a decisive factor in the tolerability outcome, not
only the criticality of individual devices. As shown in \Cref{fig:experiment-4a,fig:experiment-4b}, a non-critical device appended to an already-degraded
critical one can push $\psi$ across the baseline threshold
if the degradation rate is sufficiently high. This sensitivity
to cumulative attack composition has practical implications
for mission planning: the system's resilience boundary is
not a static property of the hardware but depends on the
current parameter configuration and the evolving content
of $S$ at each evaluation step.

%% file: chapters/ethics.tex

\section*{Ethics and Privacy Statement}
\label{sec:ethics}


All experiments were simulation-only (Webots, virtual \gls{pr2}), involving no human subjects or real systems. Data and configurations are synthetic and openly released.

%% file: chapters/genAI_usage.tex
\section*{Declaration on the Use of Generative AI}

The authors used Claude Code (Anthropic, USA) and Microsoft Copilot to help convert the \gls{pr2} controller from C to Python, automate experiments, and draft code and text. All AI-assisted output was reviewed and verified by the authors, who remain responsible for the results presented in this paper.

%% file: chapters/acks.tex




\begin{acks}
This work was supported by the Wallenberg AI, Autonomous Systems and Software Program (WASP), funded by the Knut and Alice Wallenberg Foundation.
\end{acks}